\documentclass[11pt,a4paper]{article}
\pdfoutput=1 
\usepackage{jheppub}	
\usepackage[utf8]{inputenc}
\usepackage[english]{babel}
\usepackage{makeidx}
\usepackage{tikz}
\tikzset{every picture/.style={line width=0.75pt}} 
\usepackage{amsfonts}
\usepackage{enumerate}
\usepackage{mathrsfs}
\usepackage{tensor}
\usepackage[autostyle]{csquotes}
\usepackage{subfig}

\def\be{\begin{equation}}
\def\ee{\end{equation}}
\def\bea{\begin{eqnarray}}
\def\eea{\end{eqnarray}}
\newcommand{\nn}{\nonumber}

\newcommand{\ft}[2]{{\textstyle\frac{#1}{#2}}}

\def\apjl{\ref@jnl{ApJ}}

\title{More on the SW-QNM correspondence}
\author[a]{Massimo Bianchi,}
\author[a]{Dario Consoli,}
\author[a]{Alfredo Grillo,}
\author[a]{Jos\`e Francisco Morales.}
\emailAdd{massimo.bianchi@roma2.infn.it}
\emailAdd{dario.consoli@roma2.infn.it}
\emailAdd{alfredo.grillo@roma2.infn.it}
\emailAdd{morales@roma2.infn.it}

\affiliation[a]{Dipartimento di Fisica, Università di Roma ``Tor Vergata"  \& Sezione INFN Roma2, Via della ricerca 
scientifica 1, 00133, Roma, Italy}

\abstract{  
We exploit the recently proposed correspondence between gravitational perturbations and quantum Seiberg-Witten curves to compute the spectrum of quasi-normal modes of asymptotically flat Kerr Newman black holes and establish detailed gauge/gravity dictionaries for a large class of black holes, D-branes and fuzzballs in diverse dimensions. QNM frequencies obtained from the quantum periods of $SU(2)$ 
${\cal N}=2$ SYM with $N_f=3$ flavours are compared against numerical results, WKB (eikonal) approximation and geodetic motion showing remarkable agreement. Starting from the master example relating quasi-normal modes of Kerr-Newman black holes in AdS$_4$ to $SU(2)$ gauge theory with $N_f=4$, we illustrate the procedure for  some simple toy-models that allow analytic solutions. We also argue that the AGT version of the gauge/gravity correspondence may give precious hints as to the physical/geometric origin of the quasi-normal modes/Seiberg-Witten connection and further elucidate interesting properties (such as tidal Love numbers and grey-body factors) that can help discriminating black holes from fuzzballs.
}

\preprint{PREPRINT}
 \clearpage

\begin{document}
\maketitle
\flushbottom
\section{Introduction}
\label{Intro}

Compact gravitating objects, such as black holes (BHs), D-branes and micro-state geometries (`fuzz-balls') are often characterised by a set of Quasi-Normal Modes (QNMs) \cite{Bianchi:2021xpr} that govern the linear response to external perturbations.  In the eikonal (WKB) approximation, the complex QNM frequencies can be written as 
\begin{equation}
\label{QNMWKB}
 \omega_{QNM}\approx   \omega_c(\ell)  -i (2n+1)\lambda
\end{equation}
with $\omega_c(\ell) $ the frequencies  of the (unstable) `circular' orbits forming the so-called photon-sphere, $\lambda$ the Lyapunov exponent, encoding the damping time of the wave and quantifying the chaotic behaviour of geodesics near the photon-sphere, and $n$ the so-called `overtone' number \cite{Cardoso:2008bp,Mashhoon:1985cya,Schutz:1985km,Iyer:1986np,Yang:2012he,Bianchi:2020des,Bianchi:2020yzr,ToVSapQNM}.

The interest in accurate values of $\omega_{QNM}$ is two-fold. On the one hand they dominate the Gravitational-Wave ring-down signal in binary mergers and may help discriminating BHs from  fuzzballs or other Exotic Compact Objects \cite{Bianchi:2020bxa,Bianchi:2020miz,Bena:2020see,Bena:2020uup,Bah:2021jno}. On the other hand, due to the choice of boundary conditions, QNMs solve non self-adjoint spectral problems, such as Regge-Wheeler-Zerilli or Teukolsky equations \cite{RegWheel,Zerilli:1970se,Teukolsky:1972my}, and form an over-complete set that play a crucial role in the study of BH perturbations. Alas even for the simplest (spherically symmetric) case (e.g. Schwarzschild BHs) accurate values of $\omega_{QNM}$ can only be computed via numerical methods \cite{Leaver:1985ax,Leaver:1990zz}.

Quite recently, attempting exact WKB quantization techniques \cite{Mironov:2009uv,Zenkevich:2011zx,Bourgine:2017aqk,Fioravanti:2019vxi,Grassi:2018bci,Grassi:2019coc}, a new astonishing gauge-gravity connection between the QNM spectral problem and quantum Seiberg-Witten (SW) curves \cite{Seiberg:1994rs,Seiberg:1994aj,Nekrasov:2009rc} for $\mathcal{N}=2$ SYM theories was suggested and tested in the case of Kerr BHs in 4-d \cite{Grassi}. It is worth mentioning that similar problems have been studied using the monodromy properties of differential equations yielding similar results \cite{Novaes:2014lha, CarneirodaCunha:2015hzd, CarneirodaCunha:2015qln,Amado:2017kao,Lencses:2017dgf,BarraganAmado:2018zpa,Novaes:2018fry,CarneirodaCunha:2019tia,Amado:2020zsr,Bershtein:2021uts,BarraganAmado:2021uyw,Cavalcante:2021scq,daCunha:2021jkm,Amado:2021erf}.

The QNM-SW correspondence was extended in \cite{Bianchi:2021xpr} to several gravity systems including BHs in higher dimensions, D-branes, their bound-states and fuzzballs (smooth horizonless micro-state geometries). Moreover, exploiting the AGT correspondence \cite{Alday:2009aq}, wave functions for Kerr BHs were related to correlators in two-dimensional CFTs involving degenerate fields, thus providing a new tool to study other interesting observables of the gravity solution such as Love numbers, absorption coefficients and grey body factors \cite{BonTanzetc}.

Aim of this paper is to apply these ideas to a large class of gravity backgrounds and to develop some numerical, WKB, and geodetic motion methods that allow to test the QNM results.     
QNMs are obtained as solutions of the wave equation with outgoing boundary conditions outside the photon-sphere and ingoing in the interior. For a geometry with enough isometries, the equation can be separated into ordinary Schr\"odinger like differential equations describing the wave propagation in the radial and angular directions. The  equations can be put in the canonical form {\it viz.}
\begin{equation}\label{caneq}
  {d^2\Psi \over dz^2} + Q(z) \Psi = 0
\end{equation}
with $Q(z)$ a rational function.  We find that the $Q$-functions characterising many BH and brane solutions in various dimensions can assume the form
\be
Q(z)={ P_{2n+2}(z) \over \Delta_{n+3}(z)^2 }
\ee
with $P_{2n+2}(z)$ and $\Delta_{n+3}(z)$ polynomials of order $(2n+2)$ and $(n+3)$ respectively.  The same differential equation describes the dynamics of an $SU(2)^n$ linear $\mathcal{N}=2$  quiver theory in the Nekrasov-Shatashvilli (NS) $\Omega$-background given by setting $\epsilon_1=\hbar$, $\epsilon_2=0$. More precisely, the Q-function defines the quadratic differential $\phi_2(z)=Q(z) dz^2$ of the $\Omega$-deformed version of the SW \cite{Gaiotto:2009we}.
Zeroes of $P_{2n+2}(z)$ specify the positions of the branch points of the associated SW curve, while those of  $\Delta_{n+3}(z)$ encode the gauge couplings.  Identifying the two $Q$'s one can establish a dictionary between the parameters describing the gravity solution (radial/angular variable $z$, mass $\mathcal{M}$, charge $\mathcal{Q}$, angular momentum $\mathcal{J}$, frequency $\omega$ and conserved `quantum'  numbers $\ell$, $m$'s) and the gauge theory parameters ($z_{SW}$, the RG scale $\Lambda$, hypermultiplet masses $m_f$ and Coulomb branch moduli $u_a$).

Quite remarkably, a large class of BHs and brane systems can be described in terms of SW geometries  associated to $\mathcal{N}=2$ SYM with a single $SU(2)$ gauge group, {\it i.e.} $n=1$, and $N_f$ hypers in the fundamental (doublet) representation.  For instance, the spectral problems of  AdS Kerr-Newman (KN) BHs in four dimensions lead to Heun equations with four regular singularities that can be mapped to $SU(2)$ gauge theory with four fundamentals.  The asymptotically flat KN BH non-extremal and extremal solutions arise from the general case after confluence of one or two pairs of singularities and lead to $SU(2)$ gauge theories with $N_f=3$ and $N_f=2$ fundamentals, respectively.
 
The common feature of all the solutions is the presence of a photon-sphere (or a photon-halo in the rotating case), associated to degenerate choices of the frequencies where two zeroes of $Q(z)$ coincide. In the gauge theory picture this corresponds to points in the moduli space where two branch points collide and the elliptic geometry degenerates in the absence of a NS $\Omega$-background.  The singularity is smoothed out by quantum corrections once $\epsilon_1=\hbar$ is turned on. QNMs are associated to solutions of the exact WKB quantization condition  
\begin{equation}
a_\gamma=\oint_\gamma \lambda = (n_\gamma +\nu) \hbar
\end{equation}
with $\gamma$ the degenerating cycle in the classical limit and $n_\gamma$ an integer and $\nu=0,\ft12$ depending on $\gamma$. For example, radial and angular equations will be associated to degenerations of $a_D$ and $a$-cycles respectively, with $-n_r$ parametrizing the overtone and $n_\theta=\ell$ the orbital number. We find that $\nu=0$ and $\nu=1/2$ for quantization of the $a_D$ and $a$ cycles respectively. 
The period $a_\gamma$ admits an `instanton' expansion (in powers of $\Lambda$) up to one-loop terms ($\log \Lambda$) as well as a 'semi-classical' expansion (in powers of ${\hbar}$). 

Aim of the present paper is to exploit the new gauge-gravity connection to compute the QNMs of asymptotically flat KN BHs in 4-d. In order to test our results we compare them with the numerical results obtained via continuous fractions \`a la Leaver \cite{Leaver:1985ax,Leaver:1990zz}, geodetic motion and WKB approximation, finding reassuring agreement. We will also establish detailed gauge/gravity dictionaries for several gravity solutions including  AdS KN BHs in 4-d, D3-branes and their bound-states \cite{Cvetic:1995uj}, CCLP solutions of Einstein-Maxwell gravity in 5-d \cite{CCLP1,CCLP2}, circular D1-D5 fuzz-balls \cite{Lunin:2001fv} and regular JMaRT solutions in 6-d \cite{JMaRT,GMS1,GMS2}.

We will mostly focus on massless minimally-coupled scalar perturbations\footnote{Generalization to vector and tensor modes is straightforward, though tedious and does not add much to the general features of the QNMs.}.
Extending our analysis to generic micro-state geometries, in the spirit of the fuzzball proposal, seems hard  due to the lack of  isometries that prevents from writing down (de-)coupled ODE's for the QNMs.
The use of the AGT correspondence, on the other hand, may provide additional information in discriminating BHs from fuzzballs. We intend to explore these observables for D-branes and fuzzballs in the near future.

The plan of the paper is as follows. In Section \ref{Kerr_Newman_AdS} we study QNM solutions of AdS KN scalar wave equations using semi-classical methods based on WKB and geodetic motion. In Section \ref{SW_quantization}, we review the quantum SW geometry and establish the QNM-SW correspondence for the master AdS KN example. We illustrate the algorithm in a handful of toy models and examples where analytic solutions for QNMs can be found and related to quantum periods of free gauge theories. 
The AGT version of the QNM-SW correspondence introduced in \cite{BonTanzetc} is briefly reviewed and extended to the case of $SU(2)$ gauge theory with $N_f=4$ flavours. In Section \ref{numerical_analysis} we introduce a numerical method based on Leaver's continuous fraction approximation that accounts also for extremal cases. In Section \ref{Kerr_Newman} we compare the results for QNM frequencies of KN BHs obtained via WKB, SW and numerical methods. Section \ref{other_geometries} contains the detailed QNM-SW dictionary for various BHs and brane systems, including the prototypical D3-brane discussed in \cite{Bianchi:2021xpr}. In Section \ref{conclusions} we draw some conclusions. We relegate some technical details and some tables and plots of results to Appendices \ref{appendixA}, \ref{appendixB} and \ref{appendixTables}. 
 
\section{The AdS Kerr-Newman solution: wave equation vs geodetic motion}
\label{Kerr_Newman_AdS}
We consider gravity solutions  surrounded by photon-spheres. QNMs in these geometries can be defined as solutions of the wave equation with outgoing boundary conditions at infinity and ingoing boundary conditions at the horizon or, for smooth horizonless geometries, regularity in the interior of the photon-sphere.  For concreteness we focus on massless scalar perturbations but higher spin ({\it viz.} vector and tensor) perturbations of the geometry can be studied with similar techniques.  

 In this section we consider the case of KN (charged rotating) BH solution in AdS$_4$ and derive semi-classical formulae for the QNMs  
using WKB methods and geodetic motion (see \cite{Suzuki:1998vy,Zhidenko:2003wq,Giammatteo:2005vu} for previous studies of QNMs in Kerr-AdS spacetimes).

\subsection{The wave equation }

The KN-AdS BH solution is characterised by the mass $\mathcal{M}$, the angular momentum parameter $a_{_\mathcal{J}}=\mathcal{J}/\mathcal{M}$, the electrical charge ${\cal Q} $ and the AdS size $L$. 
The line element in Boyer-Lindquist coordinates reads \cite{Caldarelli:1999xj}
\begin{equation}
\label{metric_AdS_KN}
ds^2 = -\frac{\Delta_r[dt-a_{_\mathcal{J}}\,d\phi (1-\chi ^2)]^2}{\alpha_L^2 \rho^2}
+\frac{\Delta_\chi [a_{_\mathcal{J}}\,dt-d\phi (a_{_\mathcal{J}}^2+r^2)]^2}{\alpha_L ^2 \rho ^2}
+\rho ^2 \left(\frac{dr^2}{\Delta _r}+\frac{d\chi^2}{\Delta _{\chi }}\right)
\end{equation}
where $\chi = \cos \theta$ and 
\begin{equation}
\begin{aligned}
\Delta_r &= (r^2+a_{_\mathcal{J}}^2)\left(1+\frac{r^2}{L^2}\right) - 2 \mathcal{M} r + \mathcal{Q}^2 \,,
\quad
\Delta_\chi = (1-\chi^2)\left(1-\frac{a_{_\mathcal{J}}^2\chi^2}{L^2}\right)\,,
\\
\rho^2 &= r^2 + a_{_\mathcal{J}}^2 \chi^2\,,
\qquad
\alpha_L = 1-\frac{a_{_\mathcal{J}}^2}{L^2}
\end{aligned}
\end{equation}  
We are interested on QNMs arising from scalar perturbations of the metric. On AdS they are described by the wave equation
\be
(\square - M_\Phi^2) \Phi =\left[  {1\over \sqrt{g} } \partial_M \left( \sqrt{g} \, g^{MN} \partial_N \right) -M_\Phi^2 \right] \Phi=0
\label{wavebox}
\ee
with\footnote{According to holography, this massive scalar field is dual to scalar operators of conformal dimension $\Delta=1$ or $\Delta=2$, as $M_\Phi^2 L^2=\Delta(\Delta-3)$ in $AdS_4/CFT_3$.} 
\be
M_\Phi^2=-2/L^2
\label{mphi}
\ee
QNMs on asymptotically flat spaces can be obtained by sending $L\to \infty$ and are described by massless scalar waves. 

For the KN-AdS BH metric (\ref{metric_AdS_KN}), the wave equation (\ref{wavebox}) can be separated into radial and angular equations.  Denoting by $z$ the radial or angular variable, the individual equation for the radial or angular function $\phi(z)$ takes the generic form
 \be
\phi''(z)+q_{1}(z) \, \phi'(z) +q_{0}(z) \phi(z)=0 \label{geneq}
\ee
By writing 
\be
\phi(z)=e^{-{1\over 2} \int^z q_{1}(z') dz' }  \Psi(z)
\ee
one can bring (\ref{geneq}) to the canonical form
(\ref{caneq})  
 with
\be
Q (z) = q_{0}(z) -{q_{1}(z)^2\over 4} -{q_{1}'(z)\over 2}   \label{qw}
\ee
 Explicitly, taking
\be
\Phi(t,r,\chi,\phi)=e^{{-\rm i}(\omega t-m_\phi \phi)}  \frac{R(r)S(\chi)}{\sqrt{\Delta_r \Delta_\chi}}
\ee
the wave equation separates into two equations of type (\ref{caneq}) with
\begin{equation}
\begin{aligned}
 Q_{r}  & = 
 \frac{1}{\Delta _r^2}
 \left[\alpha_L^2 \left(\omega (a_{_\mathcal{J}}^2{+}r^2){-}a_{_\mathcal{J}} m_\phi\right)^2
      -\Delta _r (K^2{+}r^2 M_\Phi^2)
      {-}\ft12 \Delta _r \Delta _r''{+}\ft14 \Delta_r'{}^2\right]  \\
  Q_{\chi} &=\frac{1}{\Delta _{\chi }^2}
  \left[{-}\alpha_L^2 \left(a_{_\mathcal{J}} \omega (1{-}\chi^2) {-} m_\phi\right)^2
        {+}\Delta_\chi (K^2 {-}a_{_\mathcal{J}}^2 \chi ^2 M_\Phi^2)
        {-} \ft12 \Delta _{\chi } \Delta_\chi''{+}\ft14 \Delta_\chi'{}^2\right]
\label{qs}
\end{aligned}
\end{equation}
and $K^2$ a separation constant. It is easy to check that for  $M_\Phi$ given by  (\ref{mphi}) the $r^6$ terms in the numerator of (\ref{qs}) exactly cancel, so that the numerator of  both the radial $Q_{r} (r)$  and angular $Q_{\chi} (\chi)$  are given by polynomials of order four. 
The resulting $Q$-characteristic functions will be put in correspondence with that describing the dynamics of $SU(2)$ gauge theory with $N_f=(2,2)$ fundamentals.  
 
Finally, the radial wave function $R(r)$ should be supplemented with the boundary conditions
\begin{equation}
R(r) \underset{r\to\infty}{\sim} e^{{\rm i}\,\omega\,r}
\, , \quad
R(r) \underset{r\to r_H}{\sim} e^{-{\rm i}\,\omega\,(r-r_H)}
\end{equation}
with $\omega=-P_t$ the frequency of the wave and $r_H$  the horizon.  On the other hand, the angular wave-function $S(\chi)$ satisfies  boundary conditions arising from periodicity and regularity at $\chi=\pm 1$.

\subsection{WKB approximation}

With the AdS-KN case in mind, let us consider the individual wave equations for radial and angular dynamics
\be
\Psi''(z) +Q (z)\,\Psi(z)=0 \label{secondD}
\ee
that can be both viewed as the Schr\"odinger equation for a particle subject to a potential $V$ with $Q=E-V$ and $E$ the energy. 
In the limit of large frequencies the equation can be solved in a semiclassical approximation by writing
\begin{equation}
\Psi(z) ={ 1 \over  \sqrt{\varphi'(z) }   } \left( C_1  e^{{\rm i}  \varphi(z)  } +C_2 \,  e^{-{\rm i} \varphi(z)  }  \right)
\end{equation}
with
\be
\varphi(z) =   \int^z \sqrt{Q(z')} ~ dz'  \label{phiz}
\ee
The approximation breaks down near the zeroes $z_\pm$ (inversion points) of  $Q(z)$ where $\varphi'(z)$ vanishes. The matching between the solutions on the two sides of the inversion points requires that the frequency $\omega$ satisfy the Bohr-Sommerfeld quantization condition
\be
\int_{z_-}^{z_+}   \sqrt{Q(z)} \,  dz={\rm } \pi \left( n+ \ft12 \right)   \quad .\label{bohr}
\ee
with $n$ a non-negative integer.
In the semiclassical limit where inversion points collide the integral can be approximated as
\begin{equation}
\int_{z_-}^{z_+} \sqrt{Q(z)} dz \approx \int_{z_-}^{z_+} \sqrt{Q(z_c) +{Q''(z_c)\over 2}(z-z_c)^2}\, dz \approx  {{\rm i} \pi Q(z_c) \over \sqrt{2 Q''(z_c) }} \label{bohr2}
\end{equation}
where $z_c\in \left[ z_-,z_+\right]$ is the extremum inside the integration contour, {\it i.e.} $Q'(z_c)=0$. 
Applying  (\ref{bohr2}) to the radial and angular equations, one finds   
\bea
Q_s'(z_c^s) &=& 0 \nn\\
{Q_s(z_c^s) \over \sqrt{2 Q_s''(z_c^s) }} &=& - {\rm i}  \left( n_s+ \ft12 \right)  \label{wkb-radial}
\eea
with $s=r,\theta$.
  To make easier the comparison against standard results in the QNM literature we use $\theta$ (rather than $\chi$) variable in the WKB analysis and introduce the separation constant $A$ instead of $K^2$
 \be
K^2=A-m_\phi^2 +\alpha_L^2 (a_{_\mathcal{J}}  \omega-m_\phi)^2
\ee
   The equation $Q_\theta'(\theta_c)=0$ can be solved by taking  $\theta_c=\pi/2$. 
   The remaining equations can be solved for the critical radius $r_c$, the frequency $\omega$ and the separation constant $A$ by giving
   to   $\omega$ and $A$ small imaginary parts. More precisely, we write
\be
\omega= \omega_c +{\rm i} \omega_I \qquad, \qquad  A= A_c +{\rm i} A_I
\ee
with $|\omega_I |\ll |\omega_c |$ and $ |A_I |\ll |A_c |$ and solve equations (\ref{wkb-radial})   order by order in $\omega_I$ and $A_I$.  
To leading order, using that $Q_{r}'' (r_c) >0$  and $Q_{\theta}'' (\theta_c) <0$, one finds  
   \bea
  &&   \partial_r Q_r(r_c,\omega_c,A_c)=Q_r(r_c,\omega_c,A_c)= 0  \nn\\
&& B_\theta (\omega_c,A_c) = {Q_\theta(\omega_c,A_c) \over \sqrt{-2 Q_\theta''(\omega_c,A_c) }}= n_\theta+ \ft12 \label{eqwave}
\eea
   with all functions evaluated at $\theta_c=\pi/2$ and
  \be
  n_\theta=\ell-| m_\phi|
   \ee
  Equations (\ref{eqwave}) can be solved for $r_c$, $\omega_c$ and $A_c$.  The imaginary parts follows from the expansion of 
  (\ref{wkb-radial}) to linear order in  $\omega_I$, $A_I$. One finds
\bea
\omega_I =- {\left( n_r+ \ft12 \right)\sqrt{2 Q_r'' }\over \partial_\omega Q_r  -\partial_A Q_r   { \partial_\omega B_\theta \over \partial_A B_\theta } }\qquad ,\qquad 
A_I = - \omega_I  { \partial_\omega B_\theta \over   \partial_A B_\theta} \label{wkim}
\eea
with  all functions evaluated at  $\omega_c$, $A_c$, $r_c$ and $\theta_c$. 

\subsection{Geodetic motion}
\label{geo_approx}
QNM frequencies in the semi-classical approximation can be alternatively derived from the geodetic motion of massless particles near  the photon-sphere of the gravitating object. In the Hamiltonian formalism, geodetic motion is described by the Hamilton-Jacobi equations
\bea
\dot{x}^M = {\partial {\cal H} \over \partial P_M} \qquad , \qquad \dot{P}_M =-{\partial {\cal H}  \over \partial x^M }
\eea
with 
\begin{equation}
\mathcal{H} =\ft12 g^{MN} P_M P_N = 0 
\end{equation}
For the AdS KN metric, the Hamiltonian can be written in the separable form
\be
2{\cal H} = \Delta_r (P_r^2-Q_{\rm r,geo}) + \Delta_\theta (P_\theta^2-Q_{\theta,{\rm geo}})
\ee
with 
\begin{align}
 Q_{r,{\rm geo} } (r) & =
 \frac{\alpha _L^2 [\omega (r^2+a_{_{\cal J}}^2)-a_{_{\cal J}} m_\phi]^2
       -\Delta _r [A{\,-\,}m_\phi^2 {\,+\,}\alpha_L^2 (a_{_\mathcal{J}}  \omega{\,-\,}m_\phi)^2  ]}
 {\Delta _r^2} \label{qs2}\\
 Q_{\theta, \rm geo} (\theta)&=
 \frac{\Delta_\theta\, \sin^2\theta\, [A{\,-\,}m_\phi^2 {\,+\,}\alpha_L^2 (a_{_\mathcal{J}}  \omega{\,-\,}m_\phi)^2]
        - \alpha_L^2 (m_\phi{\,-\,} a_{_{\cal J}} \omega \sin^2\theta )^2}
 { \Delta_\theta^2} \nn
\end{align}
The null equation ${\cal H}=0$ reduces to the one-dimensional conditions
\be
P_r(r)^2-Q_{\rm r,geo}(r)=P_\theta (\theta)^2-Q_{\rm \theta,geo}(\theta)=0 \label{pqgeo}
\ee 
 The characteristic $Q$-functions  match those in the wave equations in the eikonal limit where the last two terms in (\ref{qs})
 can be discarded and $Q_\theta\approx \sin^2\theta \, Q_\chi(\cos\theta)$. Similarly the null condition (\ref{pqgeo})   
match radial and angular wave equations after quantization of momenta, i.e. $P_s\approx -{\rm i} \partial_s \ln \Psi_s({x}^s) $.

  Zeroes of the $Q$-functions are then associated to inversion points of the geodetic motion. A double zero of $Q_r$ signals the existence of a photon sphere, i.e. null circular orbits. These geodesics are known to exist for a given range of parameters (depending on the angular momenta $K$ and $P_\phi$, the radius $r_c$ and of the frequency $\omega_c$), such that
  \be
\begin{aligned}
&Q_{\rm r,geo}(r_c,\omega_c,A_{c} )=Q_{\rm r,geo}'(r_c,\omega_c,A_{c})=0 \label{qqp2}
\end{aligned}
\ee
   These are precisely the equations defining the extremum and the real part of the QNM frequency in the WKB approximation. 
 On the other hand, the imaginary part of the frequency  can be related to the  radial velocity of a freely falling geodesics at the photon sphere\footnote{Here we used $\partial_A {\cal H}=0$ to rewrite ${\Delta_\theta\over \Delta_r}=-{ \partial_A Q_{r,\rm geo}\over \partial_AQ_{\theta,\rm geo}}$.}
\be
   {dr\over dt} =
   {{\partial{\cal H}\over\partial P_r}\over{\partial{\cal H}\over \partial P_t}}=
   {2 P_r \over \partial_\omega Q_{ r,\rm geo}{+} {\Delta_\theta\over \Delta_r} \partial_\omega Q_{\theta,\rm geo} }  =
   {2 \sqrt{Q_{r,\rm geo} } \over \partial_\omega Q_{r,\rm geo}-{\partial_A Q_{r,\rm geo} \over \partial_A Q_{\theta,\rm geo}} \partial_\omega Q_{\theta,\rm geo}}
   \approx {\,-\,} 2 \lambda (r{\,-\,}r_c) \label{lambdar}
\ee 
with
\be
\lambda =
  { \sqrt{ Q_{ r,\rm geo} ''\over 2}
  \over
  \partial_\omega Q_{ r,\rm geo} - { \partial_A Q_{r,\rm geo} \over \partial_A Q_{\theta,\rm geo}}\partial_\omega Q_{\theta,\rm geo}}
  \label{lyap}
\ee 
$\lambda$ is known as the Lyapunov exponent and quantifies the chaotic behavior of nearly critical geodesics around the photon-sphere \cite{Cardoso:2008bp,Yang:2012he,Bianchi:2020des,Bianchi:2020yzr,ToVSapQNM}. We write 
\be\label{omgeokgeo}
\omega_{\rm geo} = \omega_c-{\rm i} \lambda (2n_{\rm geo}+1)  
\ee
 that agrees with (\ref{wkim}) in the limit of large charges where  
 \be
 \partial_A Q_r   { \partial_\omega B_\theta \over \partial_A B_\theta }  \approx  \partial_A Q_{r,\rm geo} { \partial_\omega Q_{\theta,\rm geo} \over \partial_A Q_{\theta,\rm geo}}
 \ee
where all functions are understood evaluated at $\omega_,A_c,r_c,\theta_c$.

\subsection{Example: Kerr-Newman BH}

Let us illustrate the WKB formulas for the case of asymptotically flat Kerr-Newman and Schwarzschild BHs. The $Q$-characteristic functions are obtained from (\ref{qs2}) by sending the radius of AdS to infinity, i.e.  $\alpha_L \to 1$, $M_\Phi \to 0$, leading to 
\begin{equation}
\begin{aligned}
 Q_{r,{\rm geo} } (r) & =
   \frac{[\omega  (r^2+a_{_{\cal J}}^2){-}a_{_{\cal J}} m_\phi]^2}{\Delta _r^2}
  -\frac{A{-}2 m_\phi  a_{_{\cal J}} \omega {+}a_{_{\cal J}}^2 \omega^2}{\Delta _r}\\
 Q_{\theta, \rm geo} (\theta)&=
   A+ a_{_{\cal J}}^2 \omega^2 \cos^2\theta -{m_\phi^2\over \sin^2\theta}
 \label{qs2geo}
\end{aligned}
\end{equation}
  
\subsubsection*{Schwarzschild: }

  Setting $a_{_{\cal J}}=0$ and $\mathcal{Q}=0$ one finds the formulas  for the Schwarzschild BH
    \begin{equation}
\begin{aligned}
  Q_{r,{\rm geo} } ={\omega^2 r^3-A (r-2\mathcal{M}) \over r(r-2\mathcal{M})^2} \quad, \quad  Q_{\theta, \rm geo} (\theta) = A-{m_\phi^2\over \sin^2\theta} \quad, \quad 
 B_{\theta} &= { A-m_\phi^2\over 2m_\phi} \label{qs2schw}
\end{aligned}
\end{equation}
 Plugging this into (\ref{eqwave})  and (\ref{wkim}) and solving for  $\omega$, $A$, $r_c$ one finds
\be
r_c=3 \mathcal{M} \quad, \quad A=|m_\phi|(|m_\phi|{+}1{+}2 n_\theta)  \quad, \quad \mathcal{M}\, \omega_{\rm Sch}= {\sqrt{A_c} \over 3\sqrt{3}} - {\rm i} {2n_r+1\over 6 \sqrt{3} }
\ee
   
\section{Quantum Seiberg-Witten curves for \texorpdfstring{${\cal N}=2$}{N=2} SYM with flavours}
\label{SW_quantization}
In this section we first extend the QNM-SW dictionary (and its AGT version) to the general case of $SU(2)$ ${\cal N}=2$ SYM theories with $N_f=4$ hypermultiplets that will be later on associated to the wave equation of scalar metric perturbations of KN-AdS BHs. We then move on to theories with less than four hypermultiplets, which are relevant for other geometries under consideration, and show how SW quantization works in some toy examples that allows for exact solutions.
 
\subsection{The classical Seiberg-Witten curve}
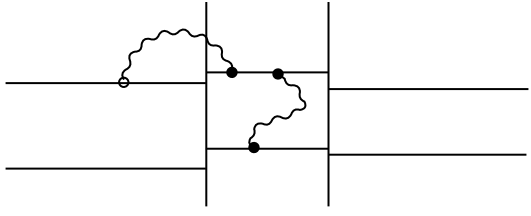
\begin{figure}[t]
\centering
\begin{tikzpicture}[x=0.75pt,y=0.75pt,yscale=-1,xscale=1]

\draw    (247,76.2) -- (247,179) ;
\draw    (308,76.2) -- (308,179) ;
\draw    (247,111.6) -- (308,111.6) ;
\draw    (246.8,150) -- (307.8,150) ;
\draw    (146.8,117) -- (246.8,117) ;
\draw    (146.8,160) -- (246.8,160) ;
\draw    (307.8,120) -- (407.8,120) ;
\draw    (307.8,153) -- (406.8,153) ;
\draw    (205.92,115.22) .. controls (204.54,113.17) and (204.86,111.49) .. (206.89,110.16) .. controls (208.96,109.21) and (209.59,107.59) .. (208.78,105.32) .. controls (208.12,103.1) and (209,101.72) .. (211.42,101.17) .. controls (213.59,101.19) and (214.69,100.04) .. (214.7,97.72) .. controls (215.05,95.31) and (216.42,94.34) .. (218.8,94.79) .. controls (220.98,95.54) and (222.5,94.85) .. (223.36,92.71) .. controls (224.55,90.61) and (226.17,90.2) .. (228.22,91.47) .. controls (230.01,92.94) and (231.69,92.81) .. (233.24,91.1) .. controls (235.09,89.54) and (236.77,89.71) .. (238.27,91.61) .. controls (239.46,93.6) and (241.09,94.07) .. (243.16,93.01) .. controls (245.23,92.1) and (246.67,92.8) .. (247.46,95.12) .. controls (247.9,97.43) and (249.21,98.4) .. (251.38,98.03) .. controls (253.89,98.16) and (255.04,99.4) .. (254.83,101.73) .. controls (254.36,103.96) and (255.24,105.34) .. (257.48,105.88) .. controls (259.75,106.79) and (260.43,108.4) .. (259.52,110.72) -- (259.8,111.6) ;
\draw [shift={(259.8,111.6)}, rotate = 73.3] [color={rgb, 255:red, 0; green, 0; blue, 0 }  ][fill={rgb, 255:red, 0; green, 0; blue, 0 }  ][line width=0.75]      (0, 0) circle [x radius= 2.34, y radius= 2.34]   ;
\draw [shift={(205.8,116.6)}, rotate = 273.69] [color={rgb, 255:red, 0; green, 0; blue, 0 }  ][line width=0.75]      (0, 0) circle [x radius= 2.34, y radius= 2.34]   ;
\draw    (270.8,149.4) .. controls (268.45,148.42) and (267.87,146.86) .. (269.06,144.72) .. controls (270.88,143.67) and (271.49,142.15) .. (270.9,140.18) .. controls (271.29,137.77) and (272.73,136.86) .. (275.22,137.44) .. controls (277.25,138.41) and (278.76,137.85) .. (279.73,135.75) .. controls (280.85,133.65) and (282.46,133.15) .. (284.56,134.26) .. controls (286.65,135.35) and (288.28,134.82) .. (289.43,132.65) .. controls (290.19,130.52) and (291.62,129.8) .. (293.71,130.49) .. controls (296.35,130.05) and (297.1,128.64) .. (295.96,126.26) .. controls (293.95,125.43) and (293.2,123.92) .. (293.71,121.73) .. controls (293.86,119.4) and (292.66,118.17) .. (290.11,118.06) .. controls (287.96,118.42) and (286.63,117.34) .. (286.12,114.81) -- (282.8,112.4) ;
\draw [shift={(282.8,112.4)}, rotate = 214.99] [color={rgb, 255:red, 0; green, 0; blue, 0 }  ][fill={rgb, 255:red, 0; green, 0; blue, 0 }  ][line width=0.75]      (0, 0) circle [x radius= 2.34, y radius= 2.34]   ;
\draw [shift={(270.8,149.4)}, rotate = 238.24] [color={rgb, 255:red, 0; green, 0; blue, 0 }  ][fill={rgb, 255:red, 0; green, 0; blue, 0 }  ][line width=0.75]      (0, 0) circle [x radius= 2.34, y radius= 2.34]   ;
\end{tikzpicture}
\caption{Brane configuration for ${\cal N}=2$ SYM with $SU(2)$ gauge group: the QFT lives on a stack of two D4-branes (horizontal lines) suspended between two non dynamical NS5-branes (vertical lines). The four external D4 flavour-branes extending to infinity provide the matter hypermultiplets of the theory.}
\label{brane_config}
\end{figure}
To get some intuition on the gauge/gravity dictionary, let us start by considering the classical SW curve for an $SU(2)$ gauge theory with $N_f= 4$ hypermultiplets with masses $m_i$ in flat space 
\be
q\, y^2  P_L(x) +y P_0(x) +P_R(x)=0
\label{swcurve}
\ee
with 
\be
P_0(x) = x^2{-}u{+}q \,p_0(x)
\, , \quad
P_R(x) = (x{-}m_1) (x{-}m_2) 
\, , \quad
P_L(x)= (x{-}m_3) (x{-}m_4) \, \,
\label{ppms}
\ee
where $q = e^{2\pi {\rm i } \tau}$ is the gauge coupling, $u = \ft12 \langle {\rm tr} \varphi^2\rangle$ the Coulomb branch
modulus and $p_0(x)$ a quadratic polynomial in $x$ determined below in \eqref{p0_expl}.
  
The SW curve \eqref{swcurve} can be derived from the brane configuration shown in Fig \ref{brane_config} \cite{Witten:1997sc}. The zeroes of $P_0(x)$ are associated to the positions of the colour D4-branes, while those of $P_{L/R}(x)$ to the positions of the flavour D4-branes. The degrees of $P_0(x)$ and $P_{L/R}(x)$ are given by the rank of the colour and flavour group respectively. Finally the distance between the NS5-branes is proportional to the square inverse gauge coupling.
\\
Solving for $y$ one finds
\be
y_\pm={1\over 2 q P_L} \left(- P_0\pm \sqrt{ P_0^2-4 q P_L P_R } \right)
\ee
The elliptic curve can be viewed as a double cover of the complex plane with branch points $e_i$ defined by
\be
P_0^2-4 q P_L P_R=\prod_{i=1}^4 (x-e_i) 
\label{discriminant}
\ee   
The  periods of the elliptic curve are defined as
\be
a=\oint_{\alpha} \lambda_0   \qquad , \qquad   a_D=\oint_{\beta}  \lambda_0 
\ee
with $\alpha$ and  $\beta$ the two fundamental cycles and
\be
\lambda_0 =\ft12(\lambda_+-\lambda_-) \qquad {\rm with} \qquad \lambda_\pm=  {1\over 2\pi {\rm i} }  x \,\partial_x \ln y_\pm(x) \, dx
\label{lambda0}
\ee
the SW differentials. The dynamics is coded in the analytic pre-potential ${\cal F}(a;q,m_i)$ in that the ${\cal N} = 2$ vector multiplet Lagrangian ${\cal L} = \int d^4\theta  {\cal F}(\Phi;q,m_i)$ and 
\begin{equation}
2\pi {\rm i}\, a_D= -{\partial {\cal F}\over \partial a}(a;q,m_i)
\end{equation}
At $q=0$ the four branch points collide in pairs at $\pm \sqrt{u}$, the $\alpha$-cycle shrinks to zero and the $a$-period becomes
\be
a \approx {1\over 2\pi{\rm i}} \oint_{\alpha}  {x P_0'(x)\over P_0(x)}\,dx =\sqrt{u} +\ldots
\ee
Alternatively, collecting powers of $x$, the curve (\ref{swcurve}) can be written as
\be
A(y)\,x^2 +  B (y)\, x +C (y) =0
\label{swcurve2}
\ee
and the periods as
\be
a= 2\int_{\tilde e_1}^{\tilde e_2} \lambda_0  \qquad , \qquad   a_D= 2 \int_{\tilde e_2}^{\tilde e_3} \lambda_0
\label{unclear_1}
\ee
with $\tilde e_i$ the zeroes of $B^2{\,-\,}4AC$ and 
\be
\lambda_0 ={ x_+-x_- \over 4\pi {\rm i} } \,  {dy \over y }   ={  \sqrt{B^2-4AC} \over 4\pi {\rm i} A y } \,  dy  
\label{lambda02}
\ee
where $x_\pm$ are the solutions of \eqref{swcurve2}.
 \subsection{The quantum curve}
In the presence of a non-trivial $\Omega$-background, $\epsilon_1=\hbar$, $\epsilon_2=0$, the dynamics of the gauge theory is described by a quantum curve obtained from the classical one after replacing $x$, $y$ by operators satisfying the commutation relation
\begin{equation}
[\hat{x},\ln \hat{y}]=\hbar
\label{commutation}
\end{equation}
The quantum curve follows from  (\ref{swcurve}) by distributing the powers of $y$ symmetrically \cite{Zenkevich:2011zx}
\be
\left[ q\,  \hat{y}^{1\over 2} \, P_L(\hat{x}) \,  \hat{y}^{1\over 2} +P_0(\hat{x})+\, \hat{y}^{-{1\over 2}} \, P_R(\hat{x}) \, \hat{y}^{-{1\over 2} }  \right] U=0
\label{diffsw}
\ee
with $P_0$, $P_L$, $P_R$ given in (\ref{ppms}) and  
\be
p_0(x) = x^2{-}(x{+}\tfrac{\hbar}{2}) \sum_i m_i  + u+  \sum_{i<j} m_i m_j+\tfrac{\hbar^2}{2}
\label{p0_expl}
\ee	 
This function is determined by requiring that the quantum SW differential, defined later in (\ref{sw_differential}), behaves at large $x$ as \cite{Poghossian:2010pn,Fucito:2011pn} 
\be
\lambda_+=\sum_{n=0}^\infty {\langle {\rm tr} \varphi^n \rangle \over x^n} =2+\frac{2u}{x^2}+\ldots 
\ee
Using $ \hat{x} \hat{y} = \hat{y} (\hat{x}+\hbar) $ to bring all the dependence on $\hat{y}$ to the left, and setting $\hat x=\hbar\, y\, \partial_y$, one can view (\ref{diffsw}) as an ordinary differential equation in the $y$-variable  
\be
\left[ q y^2 P_L( \hat{x}{+}\ft{\hbar}{2})   + y P_0(\hat{x} )+  P_R(\hat x{-}\ft{\hbar}{2} ) \right]U(y) = \left[ A(y) \hat{x}^2 +  B (y)\,  \hat{x} +C (y) \right] U(y) = 0
\label{swcurveABC}
\ee
with
\bea
A &=&  (1{+}y)(1{+} q y) \qquad , \qquad
B={-}m_1{-}m_2-\hbar +q y\left[ y(\hbar {-}m_3{-}m_4){-} \sum_i m_i \right]\\
C &=& (m_1{+}\ft{\hbar}{2})( m_2{+}\ft{\hbar}{2}) {-}u y
{+}q y\Bigg[ u+\sum_{i<j} m_i m_j {-}\ft{\hbar}{2} \sum_i m_i  {+}\ft{\hbar^2}{2} {+}  y \left( m_3{-}\ft{\hbar}{2} \right)  \left( m_4{-}\ft{\hbar}{2} \right) \Bigg]
 \nn
\eea
The differential equation \eqref{swcurveABC} can be cast in canonical from by taking
\begin{equation}
U(y)= {1\over \sqrt{y} } \, e^{-\frac{1}{2 \hbar}\,\int^y \frac{B(y')}{ y' A(y')}\, dy'}  \Psi(y)
\end{equation}
with
\be
Q_{\rm SW}(y)  = {4 \,C \,A {-} B^2{+}2\, \hbar\, y(B\, A'{-}A\, B'){+}\hbar^2\, A^2  \over  4\, \hbar^2\, y^2\, A^2 }
\ee
\subsection{The gauge/gravity dictionary}
In this section we establish the gauge/gravity dictionary using as a working example the AdS KN metric. To this aim, we first write $Q_{\rm SW}(y)$ in the form\footnote{At infinity $Q\simeq \delta_4/y^2$.}
\be
Q_{\rm SW}(y)  = \sum_{i=1}^3  \frac{\delta_i}{(y-y_i)^2}+ {\nu_1+q y (\delta_4{-}\delta_1{-}\delta_2{-}\delta_3)\over y(1+y)(1+y q)}
\label{qsw}
\ee
where $y_i=\{0,-1,{\,-\,}1/q \}$, 
\bea
{{\delta}}_1 &=& \frac{1}{4}-\frac{(m_1-m_2)^2 }{4 \hbar^2} \quad, \quad
{{\delta}}_2= \frac{1}{4}-\frac{(m_1+m_2)^2 }{4 \hbar^2}\nn\\
{{\delta}}_3&=&  \frac{1}{4}-\frac{(m_3+m_4)^2 }{4 \hbar^2} \quad, \quad
{{\delta}}_4= \frac{1}{4}-\frac{(m_3-m_4)^2 }{4 \hbar^2}
\label{masses_eqs}
\eea
and
\be
 4 \hbar^2 \nu_1= (q{-}1) (\hbar^2{+}4u){+}2 (m_1^2{+}m_2^2){+}2 q\left[ 2 m_3 m_4{+}(m_1{+}m_2)(m_3{+}m_4) {-} \hbar \sum_i m_i  \right]
\ee
To compare with gravity, one can consider an arbitrary change of variables $y \to y(z)$. The differential equation in the $z$-variable can be brought again to canonical form with the new characteristic function given by 
\be
Q(z)=Q_{\rm SW}(y) \, y'(z)^2 +{y'''(z) \over 2 y'(z) } -\frac{3}{4} \left[ { y''(z) \over  y'(z) }
\right]^2
\label{qqschwarzian}
\ee
For example, performing an $SL(2,\mathbb{C})$ transformation
\be
y={z_{24}\over z_{12}}  {z-z_1\over z-z_4}    \qquad , \qquad q=  {z_{12}z_{34}\over z_{24}z_{13} }   \label{sw1}
\ee
that maps the points $(0,-1,-1/q,\infty)$ to arbitrary points $z_i$,  one finds
\be
Q(z) ={P_4(z)\over \Delta_4(z)^2}
\ee
that corresponds to a Heun equation with four regular singularities. This matches precisely the form of the radial and angular wave equations (\ref{qs}) for AdS KN BH provided
\be
q=  {z_{12}z_{34}\over z_{24}z_{13} }\qquad , \qquad   \delta_i = {P_4(z_i)\over \Delta_4'(z_i)^2 }  \qquad , \qquad  {{\nu}}_1 
=\frac{z_{12}z_{14}}{z_{24}}\underset{z=z_1}{\rm Res} \,Q(z)  \label{sw3}
\ee
where left and right hand sides of the equations are given in terms of gauge and gravity variables respectively. In particular the $\delta_i$ determine the masses $m_i$,  while $\nu_1$ expresses $u$ in terms of gravity variables.      
\subsection{Quantum periods and exact quantization}
\label{quantum_periods}
The quantum periods $(a,a_D)$ can be computed exactly in $\hbar$ and perturbatively in $q$ using localization \cite{Nekrasov:2002qd}. Alternatively they can be derived by solving the difference equation following from the quantum SW (\ref{diffsw}) curve in the $x$ rather than in the $ y$-plane, \textit{i.e.} setting $\hat y=e^{-\hbar \partial_x}$ \cite{Poghossian:2010pn,Fucito:2011pn}
\begin{equation}
\left[ q\, P_L(x-\tfrac{\hbar}{2}) \, \hat{y} +P_0(x)+\, P_R(x+\tfrac{\hbar}{2}) \, \hat{y}^{-1}  \right] \widetilde U (x)=0
\label{dswcurve00}
\end{equation}
Introducing the functions
\be
W(x) ={1\over P_R(x+\ft{\hbar}{2}) }{ \widetilde U (x)  \over \widetilde U (x+\hbar) } 
\quad, \quad
M(x)=P_L(x-\ft{\hbar}{2})P_R(x-\ft{\hbar}{2})
\label{WMdef}
\ee
equation \eqref{dswcurve00} can be cast in the form
\be
q\, M(x) \, W(x) W(x-\hbar)  + P_0(x) W(x)  +1  =0
\label{dswcurve0}
\ee
That can be recursively solved order by order in $q$. In the small $q$ limit one can write $W(x)$ as a continuous fraction
\be
W(x) = -{1 \over P_0(x) +q\, M(x)  W(x-\hbar) } =
-{1 \over P_0(x) -{ q\, M (x)  \over P_0(x-\hbar) -{ q\, M(x-\hbar)  \over P_0(x-2\hbar)-\ldots}} }
\label{difference}
\ee
so that
\be
W(x)=  -{1 \over P_0(x) } \left(1  + { q\,  M(x)   \over P_0(x) P_0(x-\hbar)} +O(q^2)  \right)
\ee
It is easy to see that to order $q^k$, the function $W(x)$ has poles at points $x_n^+=\sqrt{u}+n \hbar $ or $x_n^-=-\sqrt{u}-n \hbar$, with $0\leq n\leq k$. The quantum period $a(u)$ can therefore be written as a sum over residues
\be
a(u) = \oint_{\alpha}  \lambda_{+}   = 2\pi {\rm i} \sum_{n=0}^\infty {\rm Res}_{\sqrt{u}+n \hbar} \lambda_{+x}(x) 
\label{aperiod}
\ee
of the $\hbar$-deformed SW differential
\be
\lambda_+(x)=-{x\over 2\pi {\rm i} } d \ln  W(x)   
\label{sw_differential}
\ee
that coincides with \eqref{lambda0} at $\hbar{\, = \,}0$. Inverting (\ref{aperiod}) one  finds $u(a)$ order by order in $q$. 
The Nekrasov-Shatashvili (NS) prepotential 
\be
{\cal F}_{NS}(a, q, m_i, \hbar) = \lim_{(\epsilon_1\epsilon_2) \to (\hbar,0)}  \epsilon_1 \epsilon_2 \log Z_{\rm Nekrasov} (a,\epsilon_1,\epsilon_2,q,m_i)
\ee
is then obtained  from the quantum version of the Matone relation \cite{Matone:1995rx,Flume:2004rp}
\be
u=- q\, {\partial {\cal F}_{NS}(a,\hbar,q) \over \partial q}
\label{matone}
\ee
after  integrating upon $q$. The integration $q$-independent constant is obtained from the one-loop prepotential, see appendix \ref{appendixA} for details. Dropping the dependence on $m_i$ and $\hbar$ one writes
\be
{\cal F}_{NS}(a,q)= {\cal F}_{\rm tree}(a,q)+ {\cal F}_{\rm 1{-}loop}(a)+ {\cal F}_{\rm inst}(a,q)
\ee
The tree level and instanton prepotentials are obtained after integration over $q$ of $u(a)$. One finds
\bea
\label{NS_prepotential}
{\cal F}_{\rm tree} &=&  -a^2 \log  q  \nn\\
{\cal F}_{\rm inst} &=& q\Bigg[ -{4a^2{\,+\,}3 \hbar^2 \over 8} {-}  {2 m_1 m_2 m_3 m_4 \over 4a^2{+}\hbar^2}  {+} \ft{\hbar}{2} \sum_i m_i  {-} \ft12 \sum_{i<j} m_i m_j \Bigg] +O(q^2) \nn\\
{\partial {\cal F}_{\rm 1{-}loop} \over \partial a} &=& 
\hbar \log \left[
\frac{\Gamma^2 ( 1+\tfrac{2a}{\hbar} )}{\Gamma^2 ( 1-\tfrac{2a}{\hbar} )}
\prod_{i=1}^{4}
{\Gamma\left({1\over 2}{+} {m_i-a\over \hbar}  \right)  \over \Gamma\left({1\over 2}{+} {m_i+a\over \hbar}  \right)}  
\right]
\eea
 Finally the $a_D$-period is given by  
\be
a_D(u)=-{1\over 2\pi {\rm i} } {\partial {\cal F}_{NS} \over \partial a}
\label{af}
\ee
Higher order terms in the instanton expansion in powers of $q$ can be obtained systematically. 
    
QNM frequencies are obtained by imposing WKB exact quantization conditions on a specific SW quantum period
   \be
a_\gamma (u) =\oint_\gamma \lambda =\hbar\left(n_\gamma+\nu \right) \label{agamma}
\ee    
defined such that $\gamma$ shrinks to zero size at the classical level and $\nu=0,\ft12$ depending on $\gamma$. We find that $\nu=0$ and $\nu=\ft{1}{2}$ for quantization of the $a_D$ and $a$ cycles respectively. The cycle $\gamma$ is determined by using the QNM/SW dictionary to map the colliding zeroes of the characteristic function $Q_{\rm geo}(z)$ governing the geodetic motion to colliding branch points in the SW gauge theory picture.
\subsection{Theories with \texorpdfstring{$N_f<4$}{Nf lesser than 4}}
\label{quantum_periods_lesser_flavours}
Theories with less fundamentals can be found by decoupling some of the hypers by sending their mass $m\to \infty$ and $q\to 0$, keeping finite the product $\tilde{q}=-m q$ parametrizing the gauge coupling of the new theory that will be renamed $q$ for simplicity\footnote{Every time a flavour in $P_R$ decouples, the $y$ variable must be rescaled $\tilde{y} = - y/m$ in order to keep the SW curve as in eq (\ref{swcurveABC}).}. The general case will be labelled by $N_f =({N}_L,{N}_R)$\footnote{In all cases we consider $N_L,N_R\le 2$ which leads to second order differential equations.} and $N_f={N}_L + {N}_R \leq 3$.
 The curve is given as in (\ref{swcurveABC}) with
\begin{equation}
P_L(x) = \prod_{i=3}^{2+{N}_L}(x-m_i)\,,\qquad
P_R(x) = \prod_{i=1}^{{N}_R}(x-m_i)
\end{equation}
and
\begin{equation}
P_0(x) = x^2 - u + q\,\delta_{N_f,\,3} \left(x - \sum_i m_i + \frac{\hbar}{2}\right) + q\,\delta_{N_f,\,2}
\end{equation}
The correct expression for the NS prepotential can be obtained from (\ref{NS_prepotential}) performing the limit $q\to 0$, $m\to\infty$. 
For example for $N_f = (1,2)$ one finds
\begin{equation}
\begin{aligned}
\label{NS_prepotential_Nf_12}
{\cal F}_{\rm tree} &=  -a^2 \log\left(-\frac{q}{\hbar}\right) \,,
\qquad
{\cal F}_{\rm inst} = q\Bigg[ \frac{1}{2}\sum_{i=1}^3 m_i +\frac{2 m_1 m_2 m_3 }{ 4a^2{-}\hbar^2}- \frac{\hbar}{2} \Bigg] +O(q^2) 
\\
{\partial {\cal F}_{\rm 1{-}loop} \over \partial a} &=
\hbar \log \left[
\frac{\Gamma^2 ( 1+\frac{2a}{\hbar} )}{\Gamma^2 ( 1-\frac{2a}{\hbar} )}
\prod_{i=1}^{3}
{\Gamma\left({1\over 2}{+} {m_i- a\over \hbar}  \right)  \over \Gamma\left({1\over 2}{+} {m_i+a\over \hbar}  \right)}  
\right]
\end{aligned}
\end{equation}
 where the extra contributions in the tree-level component of the prepotential is produced by the one-loop term in the decoupling limit.
 Similarly the characteristic $Q$-function is given by (\ref{qsw}) with
\begin{equation}
\begin{aligned}
A &=  1{+}y \,,
\qquad
B=q y^2 +q y -m_1 - m_2 - \hbar
\\
C &= -q y^2 (m_3 - \ft{\hbar}{2})- u y + q y\Bigg[\ft{\hbar}{2}-\sum_{i=1}^3 m_i \Bigg] + (m_1 + \ft{\hbar}{2})( m_2 + \ft{\hbar}{2}) 
\end{aligned}
\end{equation}    
\subsection{Examples at \texorpdfstring{$q=0$}{q=0}}
It is instructive to illustrate the various ingredients of the gauge/gravity dictionary in the simple case of gravity backgrounds related to free gauge theories ($q=0$). Setting $q=0$ in the $N_f=(2,2)$  curve  (\ref{swcurve}) one finds 
\begin{equation}
\left[ y \, P_0(\hat x)  + P_R(\hat x -\ft{\hbar}{2} ) \right]U(y) = 0 \label{difq0}
\end{equation}
with $\hat{x}=\hbar\, y\, \partial_y$ and
\be
P_0(x)=x^2-u \qquad, \qquad P_R(x)=(x-m_1)(x-m_2)
\ee
Writing  
\be
\Psi(y)= y^{-\frac{m_1+m_2}{2 \hbar}} (1+y)^{\frac{\hbar+m_1+m_2}{2 \hbar}} ~ U(y)
\ee
one can bring the differential equation (\ref{difq0}) to canonical form with
\begin{equation}
Q_\textup{free}(y)=
\frac{\hbar^2-(m_1-m_2)^2}{4\hbar^2 y^2}+
\frac{\hbar^2-(m_1+m_2)^2}{4 \hbar^2(y+1)^2}+\frac{2 (m_1^2+m_2^2)-4 u-\hbar ^2}{4\hbar ^2 y(y+1)}
\label{Qfree}
\end{equation}
and
\be
 \nu_1=\frac{ m_1^2+m_2^2}{2 \hbar^2}-\frac{u}{\hbar^2} - \frac{1}{4}
\ee
The solutions to the canonical equation are hypergeometric functions
\begin{equation}
\begin{aligned}
\Psi(y)=&\, d_1\, y^{\frac{\hbar+m_1-m_2}{2 \hbar}} (1{\,+\,}y)^{\frac{\hbar+m_1+m_2}{2 \hbar}} \,{}_2 F_1(\tfrac{1}{2}+\tfrac{m_1-\sqrt{u}}{\hbar},\tfrac{1}{2}+\tfrac{m_1+\sqrt{u}}{\hbar},1+\tfrac{m_1-m_2}{\hbar}|-y) 
\\
+ &\, d_2 \, y^{\frac{\hbar-m_1+m_2}{2 \hbar}} (1{\,+\,}y)^{\frac{\hbar+m_1+m_2}{2 \hbar}} \,{}_2 F_1(\tfrac{1}{2}+\tfrac{m_2-\sqrt{u}}{\hbar},\tfrac{1}{2}+\tfrac{m_2+\sqrt{u}}{\hbar},1+\tfrac{m_2-m_1}{\hbar}|-y)
\end{aligned}
\end{equation}
In appendix \ref{appendixB} we show how one can recover these solutions from the difference equation \eqref{dswcurve00}.

Two special cases of analytic solutions: {\it spherical harmonics} and {\it inverted hydrogen atom} will be discussed in the following, while {\it static} and {\it super-radiant} modes of KN BHs will be discussed later on.
\subsubsection{Spherical harmonics}
Spherical harmonics $Y_{\ell m}(\theta, \varphi)$ are defined as eigenfunctions of the Laplacian on the 2-sphere
\be
\nabla^2_{S^2} Y_{\ell m}(\theta, \varphi)=-\ell(\ell+1) Y_{\ell m}(\theta, \varphi)
\ee
Writing $Y_{\ell m}(\theta, \varphi)= e^{i m \varphi} U_{\ell m}(\chi)$, the equation can be written as
\begin{equation}
\label{Laplace_equation}
\left[\frac{\partial}{\partial \chi}\left( (1-\chi^2) \frac{\partial}{\partial \chi} \right)+\left(\ell(\ell+1)-\frac{m^2}{1-\chi^2}\right)\right] {U}_{\ell m}(\chi)=0
\end{equation}
where $\chi= \cos \theta$. With generic boundary conditions this equation admits two solutions
\begin{equation}
U_{\ell m}(\chi) = c_1\,P_{\ell\,m}(\chi) + c_2\,Q_{\ell\,m}(\chi)
\end{equation}
where
\begin{equation}
\label{associated_Legendre_functions}
\begin{aligned}
P_{\ell m}(\chi) &= \frac{1}{\Gamma (1- m)}\left(\frac{1+\chi}{1-\chi }\right)^{\frac{m}{2}} {}_2F_1\left(-\ell ,\ell +1;1- m ;\frac{1-\chi }{2}\right)
\\
Q_{\ell m}(\chi) &= \frac{\pi}{2\sin\pi m}  \left[\cos\pi  m\,P_{\ell m}(\chi ) - \frac{\Gamma (\ell + m +1)}{\Gamma (\ell - m +1)}\,P_{\ell,\,-m}(\chi )\right]
\end{aligned}
\end{equation}
are Legendre  associated functions of the first and second kind respectively. Requiring regularity at $\chi=1$ and $\chi=-1$, one finds that  $c_2=0$  and the solution reduces to the Legendre associated polynomials $P_{\ell m}(\chi )$ with  $\ell$, $m$ integers and $\ell\geq |m|$.

Now let us see how this result is recovered from the WKB exact quantization of the SW period. Writing 
 \be
 {U}_{\ell m }(\chi)={1\over \sqrt{1-\chi^2}} \Psi(\chi)
 \ee
  the equation (\ref{Laplace_equation}) can be brought to the canonical form with
  \begin{equation}
Q_\chi(\chi)= \frac{(1-m^2)(1+\chi^2)}{2\left( 1{\,-\,}\chi^2 \right)^2}+
\frac{2 \ell(\ell{\,+\,}1){\,+\,}1{\,-\,}m^2}{2 (1-\chi^2) } 
\end{equation}
The gauge gravity dictionary reads\footnote{Here and below we always fix the sign ambiguities in the dictionary as will.}
\begin{equation}
q=0 \quad, \quad
\frac{u}{\hbar^2}= (\ell+\tfrac{1}{2})^2 \quad, \quad
m_1=0 \quad , \quad
\frac{m_2}{\hbar}= |m| \quad; \quad
y=-\frac{1}{2}(1-\chi)
\label{dicAng_free}
\end{equation}
At large $\ell,m$, the conditions $Q_\chi(\chi_c)=Q_\chi'(\chi_c)=0$ are solved by $\chi_c=0$ and $\ell=|m|$. Translating back into the gauge variables one finds that $a_\gamma=\sqrt{u}-m_2=a-m_2 \approx 0$ in the classical limit. The cycle $\gamma$ contains then the $a$-cycle and the pole of $\lambda$ at $x=m_2$. Turning on $\hbar$ one finds
\be
 a_\gamma  =\sqrt{u}-m_2=(\ell-|m|+\ft12) \hbar \label{nsph}
\ee
in agreement with (\ref{agamma}) for $n_\gamma=\ell-|m|$, or equivalently
\be
a =\hbar \left(\ell +\frac{1}{2}  \right)
\ee
\subsubsection{`Inverted' hydrogen atom}
Another simple toy-model admitting QNMs is the ``inverted hydrogen atom",  obtained by flipping the sign of the hydrogen atom effective potential, \textit{i.e.} the hydrogen-like potential with repulsive charges and an imaginary angular momentum
\begin{equation}
V(r)=\frac{\mu}{r}-\frac{\lambda^2+ \tfrac{1}{4}}{r^2}
\end{equation}
where $\mu$ and ${\lambda}^2$ are taken positive. 
The general solution to the differential equation 
\begin{equation}
-\frac{1}{r^2} \frac{\partial^2 }{\partial r^2} \left[ r^2 \psi(r)\right]+\left(\frac{\mu}{r}-\omega^2 -\frac{{\lambda}^2+\tfrac{1}{4}}{r^2}\right) \psi(r) = 0
\label{difid}
\end{equation}
is given by
\begin{equation}
\begin{aligned}
\psi(r)=& c_1 e^{-i \omega r} (r \omega )^{-\frac{3}{2}-i {\lambda}} \, _1F_1(\tfrac{1}{2}-i {\lambda}-\frac{i \mu }{2 \omega },1-2 i {\lambda}|2 i \omega r) +\\
&+c_2 e^{-i \omega r} (r \omega )^{-\frac{3}{2}+i {\lambda}} \, _1F_1(\tfrac{1}{2}+i {\lambda}-\frac{i \mu }{2 \omega },1+2 i {\lambda}|2 i  \omega r)
\end{aligned}
\end{equation}
We look for solutions with in-going boundary conditions at $r=0$\footnote{In this toy-model $r=0$ plays the role of the `horizon'.} and outgoing at $r=\infty$. The in-going wave requirement at $r=0$ leads to $c_2=0$ while  the  outgoing behaviour at $\infty$ boils down to the quantization condition
\begin{equation}
\frac{1}{2}-i {\lambda}+\frac{i \mu}{2 \omega}= -n
\end{equation}
or equivalently to
\begin{equation}
\omega=\frac{\mu}{2{\lambda}+2i (n+\tfrac{1}{2})} \label{winv}
\end{equation}
Now let us see how to re-derive this result from the exact SW quantization. 
First, writing $\psi=r^{-2} \Psi_{\rm grav}(r)$,  one can bring  (\ref{difid})  
to canonical form with
\begin{equation}
Q(r)=\omega^2-\frac{\mu}{r}+\frac{{\lambda}^2+\tfrac{1}{4}}{r^2} \label{qinv}
\end{equation}
This function can be mapped to the $Q$-function  of $N_f=(1,1)$ free theory  $q=0$
\begin{equation}
Q_\textup{SW}(y)=-\frac{m_1}{y^3 \hbar ^2}+\frac{\hbar ^2-4 u}{4 y^2 \hbar ^2}-\frac{1}{4 y^4 \hbar ^2}
\end{equation}
Comparing the two $Q$-functions, one finds the gauge/gravity dictionary
\begin{equation}
q =  0 \quad, \quad
\frac{u}{\hbar^2}= -{\lambda}^2 \quad, \quad
\frac{m_1}{\hbar}= -\frac{i \mu}{2 \omega} \quad, \quad
r = - \frac{i}{2 \hbar  \omega y}
\end{equation}
 To understand which cycle shrinks at the classical level, we consider the geodetic motion near the ``photon-sphere'' defined by the critical conditions
\be
Q_{\rm geo} (r_c,\omega_c)=Q_{\rm geo}' (r_c,\omega_c)=0
\ee
with
\begin{equation}
Q_{\rm geo}(r)=\omega^2-\frac{\mu}{r}+\frac{{\lambda}^2}{r^2} \label{qinvgeo}
\end{equation}
 The solution reads
\be
r_c \approx {2\lambda^2\over \mu} \, , \qquad \omega_c \approx  {\mu\over 2 \lambda}
\ee
or in the gauge theory variables $m_1 \approx -\sqrt{u}$. We conclude that the vanishing period  is $a_\gamma=\sqrt{u}+m_1=a+m_1$, i.e. the cycle 
including the $a$-cut and the poles at $m_1$. 
Turning on $\hbar$ one finds the exact WKB quantization condition
\begin{equation}
a_\gamma=\oint_\gamma \lambda_{SW}= \sqrt{u}+m_1 = \hbar (n+\tfrac{1}{2})
\end{equation}
in agreement with (\ref{winv}). 
\subsection{The AGT picture}
\label{sagt}
We would like to conclude this section with another precious tool available in the study of $\mathcal{N}=2$ SYM: the AGT correspondence \cite{Alday:2009aq}. Thanks to this correspondence, one can relate the characteristic function $Q_{SW}(y)$, the wave function $\Psi(y)$ and the gauge partition function $Z$ of $\mathcal{N}=2$ SYM to correlators of two-dimensional Conformal Field Theories (CFTs) \cite{BonTanzetc}. In this section, we summarize the main ingredients of the dictionary in view of its application to the study of QNMs and other observables of BHs, D-branes and fuzz-balls. For $SU(2)$ SYM with $N_f=4$, which underlies all the examples analysed here, one can consider a Liouville theory with background charge $Q$ and central charge $c$ given by
\be
c=1+6\,Q^2 \quad, \quad  Q=  b+{1\over b}   \quad, \quad   b=\sqrt{\epsilon_1\over \epsilon_2}
\ee
and denote by $V_{\alpha_i}=e^{2\alpha_i \phi} $ the chiral operators of dimensions 
\be
h_i=\alpha_i(Q-\alpha_i)
\ee
where $i=1,\dots, 4$, corresponding to the four flavours. According to AGT, the gauge partition function is related to the four-point function
\be
Z=e^{{\cal F} \over \epsilon_1 \epsilon_2} = \left \langle   V_{\alpha_1}(y_1) V_{\alpha_2}(y_2) V_{\alpha_3}(y_3) V_{\alpha_4}(y_4)  \right\rangle
\ee
with
\be
q={y_{12}y_{34}\over y_{24}y_{13} }   
\ee
We consider the NS limit, $\epsilon_1=\hbar$, $\epsilon_2\to 0$, leading to
\be
b\to 0 \quad, \quad \alpha_i \to \infty  \quad, \quad b\, \alpha_i = {\rm finite} \label{blimit}
\ee
The characteristic function $Q_{\rm SW}(y)$ is identified with the ratio\footnote{The coefficients $c_{2,3}$ are determined by matching the asymptotic $Q_{\rm SW}(y) \approx h_4/y^2$ at infinity, that boils down to the conditions 
\bea
\sum_{i=1}^3 c_i=0 \qquad ,\qquad  \sum_{i=1}^3 (h_i+c_i y_i)=h_4
\eea  
}
\bea
Q_{\rm SW}(y)  &=& b^2 { \left\langle T(y) \prod_{i=1}^4 V_{\alpha_i} (y_i)  \right\rangle \over \left \langle  \prod_{i=1}^4 V_{\alpha_i}(y_i)  \right\rangle} 
= b^2 \sum_{i=1}^3 \left(  {h_i \over (y-y_i)^2 } + {c_i\over y-y_i} \right) 
\eea
with $y_i=\{ 0,-1,-1/q,\infty \}$, $h_i$ the dimensions and
\be
b\, \alpha_{1,2}=\ft{1}{2}   + {m_1\mp m_2\over 2\hbar}  \qquad, \qquad b \,\alpha_{3,4}=\ft{1}{2}   + {m_3\pm m_4\over 2\hbar}  
\label{balphai}
\ee
We notice that in the double-scaling limit (\ref{blimit}) the combinations $\delta_i=b^2 h_i$ and $\nu_i=b^2 c_i$ are kept finite.
Finally the wave function is associated to the five-point correlation function \cite{BonTanzetc}
\be
\Psi(y) =  \left\langle V_{\alpha_{12} }(y) V_{\alpha_1} (y_1) \ldots    V_{\alpha_4} (y_4) \right\rangle
\ee
involving the insertion of a degenerate field with $\alpha_{12}=-\ft{b}{2}$. The function $ \Psi(y) $ satisfies the BPZ equation \cite{Belavin:1984vu}
\bea
\Psi''(y)  +b^2 \sum_{i=1}^4 \left[  { h_i \over (y-y_i)^2 } +    { \partial_{y_i}  \over (y-y_i)}    \right] \Psi(y) =0
\label{bpz}
\eea 
that follows from the fact that  ${\cal O}_{12}=\left( L_{-1}^2+b^2  L_{-2} \right) V_{\alpha_{12}}$ is a null state, and its insertion inside any correlator leads to a vanishing result. In the limit $\hbar \to 0$, $\Delta_{12} \ll \Delta_i$, so the insertion of the degenerate field modifies the correlator only slightly leading to $\partial_{y_i} \Psi(y) \approx c_i  \Psi(y)$ with $c_i$ some constants. One finds then again a differential equation in the canonical form that can be mapped  to the equation for the QNMs in gravity after proper identification of the parameters.

\section{Numerical analysis}
\label{numerical_analysis}
In section \ref{Kerr_Newman} we will compute the spectrum of QNMs for Kerr Newman black holes using geodetic motion and SW techniques. In order to test those results, in this section we apply (and extend) the method of continuous fractions introduced by Leaver in  \cite{Leaver:1985ax,Leaver:1990zz} to find numerical estimates of the frequencies for differential equations of the form (\ref{secondD}) with
\be
Q(z)={P_4(z) \over \Delta_2(z)^2} 
\ee
This will be the relevant case  for the study of QNMs of KN solutions in the section \ref{Kerr_Newman}. We will also show that the same equation describes the extremal case after a proper variable redefinition.
\subsection{Radial equation}
\label{Leaver_radial}
The radial wave equation of the KN solution shows two regular singularities\footnote{For notational simplicity we assume $z_\pm$ to be real in general. Extension to complex $z_\pm$ is straightforward.} at $z = z_\pm$ and an irregular singularity at $z = \infty$. We  look for a solution away from the singularities, i.e. for $z>z_+ > z_-$. We start from the ansatz  
\begin{equation}
\Phi(z) = e^{ \nu z}(z{-}z_-)^{\sigma_-}(z{-}z_+)^{\sigma_+}\sum_{n=0}^\infty {c}_n \left(\frac{z{-}z_+}{z{-}z_-}\right)^n \label{ansatz}
\end{equation}
The constants $\sigma_+$, $\nu$ are determined by requiring that the ansatz solves the differential equation near $z_+$ and infinity \textit{viz.}
\begin{equation}
\nu^2 = - \, {P^{(4)}_4(z_+)\over 4!} \quad,\quad  \sigma_+(\sigma_+ {-}1) {+} \frac{P_4(z_+)}{\delta} =0
\end{equation}
with $\delta=z_+ - z_-$. The boundary conditions on $z_+$ and infinity select which solution of the above conditions should be used, e.g. if $z_+$ is an horizon to get the QNMs frequencies we have to choose $\sigma_+$ such that the wave is incoming, i.e. $\text{Im }\nu>0$. On the other hand $\sigma_-$ is conveniently fixed by requiring that the recursion involves only three terms. One finds
\begin{equation}
\sigma_- = -\sigma_+ -\nu\, \delta- { P^{'''}_4(z_+)\over 12 \, \nu} 
\end{equation}
Plugging the ansatz into the wave equation one finds the recursive relation 
\begin{equation}
\begin{aligned} \label{abeq}
& \alpha_n\, {c}_{n+1} + \beta_n\, {c}_n + \gamma_n\, {c}_{n-1} = 0 
\end{aligned}
\end{equation}
with $c_{-1}=0$ and 
\begin{equation}
\begin{aligned}
\alpha _n &= -\delta(1+n)(n+2 \sigma_+)\\
\beta _n &= 2 \delta  \left[(n+\sigma _+) (n-\delta  \nu-\sigma _-)+\left(1-\sigma _+\right) \sigma _+\right] - P_4'(x_+)\\
\gamma _n&= P_4'(x_+) {-}\frac{1}{2} \delta  P_4''(x_+)
	{-}\delta \left[ \left(\delta  \nu {+}2 (\sigma _-{+}\sigma _+)\right)\nu \delta{+}(n{-}\sigma _-{-}\sigma _+) (n{-}\sigma _-{+}\sigma _+ {-}1)\right]
\end{aligned}
\end{equation}
Finally QNM frequencies $\omega_n$ associated to the overtone $n$ can be obtained by truncating the recursion to a chosen level  (taken to be large) and solving 
numerically the
equation
\begin{equation}
\label{continuous_fraction}
\beta_n = \frac{\alpha_{n-1} \, \gamma_n}{\beta_{n-1} - \frac{\alpha_{n-2}\, \gamma_{n-1}}{\beta_{n-2}-\ldots}} +\frac{\alpha_n\gamma_{n+1}}{\beta_{n+1} + \frac{\alpha_{n+1}\gamma_{n+2}}{\beta_{n+2}  \ldots}}
\end{equation}
viewed as an equation for $\omega_n$.
\subsection{Angular equation}
\label{Leaver_angular}
The angular differential equation for KN BHs has the same singularity structure as the radial one, but now we look for a regular solution in the interval $z_- \leq z \leq z_+$. To this aim it is convenient to change the ansatz \eqref{ansatz} with
\begin{equation}
\Phi(z) = e^{ \nu (z-z_-) } (z{-}z_-)^{\sigma_-}(z_+{-}z)^{\sigma_+}\sum_{n=0}^\infty c_n^\chi (z{-}z_-)^n
\end{equation}
The exponents $\sigma_\pm$ are determined by requiring that the ansatz solves the differential equation and is regular in $z_\pm$, while $\nu$ is chosen such that the recursion involves only three terms
\begin{equation}
\sigma_\pm = \frac{1}{2} + \frac{1}{2} \sqrt{1- \frac{4 P_4(z_\pm)}{\delta^2}}
\quad, \quad
\nu^2 = - \, {P^{(4)}_4(z_-)\over 4!}
\end{equation}
with coefficients
\be
\begin{aligned}
\alpha _n^\chi &= \delta^2 (1+n)(n+2 \sigma_-)\\
\beta _n^\chi &=\delta  \left[2 \left(n+\sigma _-\right) \left(1+\delta  \nu -\sigma _+ - \sigma _-\right)-n (n+1)\right]+P_4'\left(z_-\right)\\
\gamma _n^\chi &=2 \delta  \nu  \left(1+\delta  \nu -\sigma _+ - \sigma _- - n\right)-\frac{\delta}{6}  P_4^{(3)}\left(z_-\right)
\end{aligned}
\ee
The recurrence relation can be solved for the separation constant. The angular overtone number $n$ turns out to be identified with the difference $\ell-|m|$ as in (\ref{nsph}).

\subsection{Extremal geometries}

The methods in the previous sections do not apply straightforwardly to extremal geometries in $D=4$ or higher dimensions. In the extremal case, the wave equation has two irregular singular points (infinity and the horizon), and therefore the above ansatz should be modified. One can expand the solution around a regular point \cite{Onozawa:1995vu}, however this typically yields a recurrence relation with four or five terms.

Here we circumvent the problem by  mapping the $N_f=(1,1)$ gauge theory to the $N_f=(1,2)$ theory studied in the previous subsections. This can be done whenever the masses of all hypers coincide. 

The gauge theory variable $Y$ in the $N_f=(1,2)$ theory is related to $y$ in the $N_f=(1,1)$ theory with $m_3=m_1$ via
\begin{equation}
y= -q^{-\ft{1}{2}} \left(\sqrt{Y}\pm\sqrt{1+Y}\right)^2
\end{equation}
with lowercase and capital parameters identified as
\begin{equation}
Q=-4 \sqrt{q} \quad, \quad
U=u+2\left(m_1-\frac{\hbar}{2}\right)\sqrt{q}  \quad, \quad
M_1= 0 \quad, \quad
M_2=\frac{\hbar}{2} \quad, \quad
M_3= m_1 
\end{equation}
\section{QNMs of Kerr-Newman BH}
\label{Kerr_Newman}
In this section we compute the spectrum of QNM frequencies for Kerr-Newman BHs using the three methods: geodetic motion, SW and continuous fractions (numerical).  We also discuss two special cases of analytic solutions: {\it static} and {\it super-radiant} modes related to free gauge theories. The  metric and gauge/gravity dictionary for the  Kerr-Newman BH solution can be obtained from the previous AdS case by sending $L\to\infty$.

\subsection{Gauge/gravity dictionary} 
 The line element of the KN metric in Boyer-Lindquist coordinates reads \cite{Caldarelli:1999xj}  
\begin{equation}
\label{metric}
ds^2=-\frac{{\Delta_r}}{\rho ^2}(dt- a_{_\mathcal{J}}\sin^2\theta\, d\phi)^2 + \frac{\sin^2\theta}{\rho ^2}\left[a_{_\mathcal{J}}\,dt- 
(r^2{+}a_{_\mathcal{J}}^2)\,  d\phi \right]^2 + \frac{\rho ^2 dr^2}{{\Delta_r}} + \rho ^2 d\theta^2 
\end{equation}
where 
\begin{equation}
\begin{aligned}
\Delta_r &=r^2 -2 {\cal M} r +{\cal Q}^2+a_{_\mathcal{J}}^2\,,
\qquad
\rho^2 = r^2 + a_{_\mathcal{J}}^2\cos^2\theta  \,,
\end{aligned}
\end{equation}
These BHs posses two generally distinct horizons located at the zeroes of $\Delta_r$
\be
r_\pm  =  {\cal M}  {\pm} \sqrt{  {\cal M}^2{-}a_{_\mathcal{J}}^2{-}{\cal Q}^2 }
\ee
 The wave equation can be separated using the  ansatz  
\be
\Phi(t,r,\theta,\phi)=e^{{-\rm i}(\omega t-m_\phi \phi)}  \frac{R(r)S(\chi)}{\sqrt{(1-\chi^2)\Delta_r}}
\ee
with $\chi=\cos\theta$. The radial and angular $Q$-functions read
\begin{equation}
\begin{aligned}
 Q_{r}  &= \frac{\left( \omega (r^2+a_{_\mathcal{J}}^2) - a_{_\mathcal{J}} m_\phi \right)^2-\Delta _r \left( a_{_\mathcal{J}}^2 \omega^2 -2 a_{_\mathcal{J}} \omega m_\phi+A \right)+\frac{1}{4} \left(\Delta _r^{'}{}^2-2 \Delta _r^{''} \Delta _r\right)}{\Delta _r^2}
 \\
 Q_{\chi}&=  \frac{(1-\chi ^2) (a_{_\mathcal{J}}^2 \omega ^2\chi ^2+A)-m_{\phi }^2+1}{(1-\chi^2)^2}\label{qswkn}
 \end{aligned}
 \end{equation}
where $A$ is the separation constant and $m_\phi$ the azimuthal angular momentum of the incoming particle.
Both differential equations on the gravity side can be matched with that of $SU(2)$ gauge theory with $N_f=(1,2)$ fundamentals. The dictionary for the radial equation reads
\begin{equation}
\begin{aligned}
{q\over \hbar} &= 2\,{\rm i}\, \omega\, ( r_+{-}r_-)\,,
	\\
	{u\over \hbar^2}  &=A+2 m_{\phi }a_{_\mathcal{J}} \omega -\omega ^2 \left(3 a_{_\mathcal{J}}^2+4 r_+ \left(r_-+r_+\right)\right)+\left(\frac{1}{2}+i \left(r_+-r_-\right) \omega \right)^2
	 \\ 
	{m_1\over \hbar} &={m_3\over \hbar}= - {\rm i} (r_++r_-) \omega
	\quad, \quad  
	{m_2\over \hbar} = -\frac{{\rm i} \left[\left(r_-^2+r_+^2+2a_{_\mathcal{J}}^2\right) \omega -2 a_{_\mathcal{J}} m_\phi \right]}{r_+-r_-} 
	 \,; \quad
	 \\
	 y &= -{r{-}r_-\over r_+ {-} r_-}\label{dicKN}
\end{aligned} 
\end{equation}
whereas for the angular equation one has
\begin{equation}
\begin{aligned}
	{q^\chi \over \hbar} &= 4a_{_\mathcal{J}} \omega \quad, \quad 
	{u^\chi \over \hbar^2}=a_{_\mathcal{J}} \omega  \left(a_{_\mathcal{J}} \omega -2 m_{\phi }+2\right)+A+\frac{1}{4} \\
	{m_2^\chi \over \hbar} &= m_\phi \quad, \quad m^{\chi}_1=m^{\chi}_3=0 \quad , \quad y^\chi = -\frac{1-\chi}{2}
\end{aligned}
\end{equation}
The extremal limit is obtained by sending $r_- \to r_+$.  This corresponds to taking $q\to 0$ and $m_2\to \infty$ while their product $q_\text{ext}=-q\,m_2$ remains finite. The resulting theory has $N_f=(1,1)$ fundamentals with
\begin{equation}
\begin{aligned}
   {q_\text{ext} \over \hbar^2} &=   4 \, \omega \left[  {a_{_{\cal J}}}\, m_\phi-\omega\, (r_+^2+a_{_{\cal J}}^2)  \right] 
   \quad, \quad 
 	{u\over \hbar^2}  =  A +\frac{1}{4}+2 m_\phi a_{_\mathcal{J}} \omega-(3a_{_{\cal J}}^2+ 8 r_+^2 )\omega^2 \,,
 	\\
	{m_1\over \hbar} &= {m_3\over \hbar}= -2{\rm i} r_+ \omega\,
	\quad; \quad 
	y_\text{ext} = -\frac{y}{m_2} = \frac{{\rm i}}{2 \hbar}\frac{r-r_+}{(r_+^2 + a_{_{\cal J}}^2)\omega - a_{_{\cal J}} m_\phi}
	\end{aligned}
\end{equation}

\subsection{SW vs Numerical vs WKB }
To showcase the computation of QNMs using the three methods, let us  consider an explicit KN solution with
\begin{equation}
{\cal M} = 1
\quad , \quad
{\cal Q} = 0.5
\quad , \quad 
a_{_{\cal J}} = 0.3
\quad , \quad
\ell = m_\phi = 2
\end{equation}
 The simplest computation is given by the eikonal WKB or geodetic motion.  Real parts of the QNM frequency and separation constants 
 are given by solving (\ref{eqwave}) with $Q_{\rm geo}$ given by (\ref{qs2geo}). Imaginary parts follow from (\ref{wkim}). 

\begin{figure}[t]
\center
\includegraphics[scale=0.353]{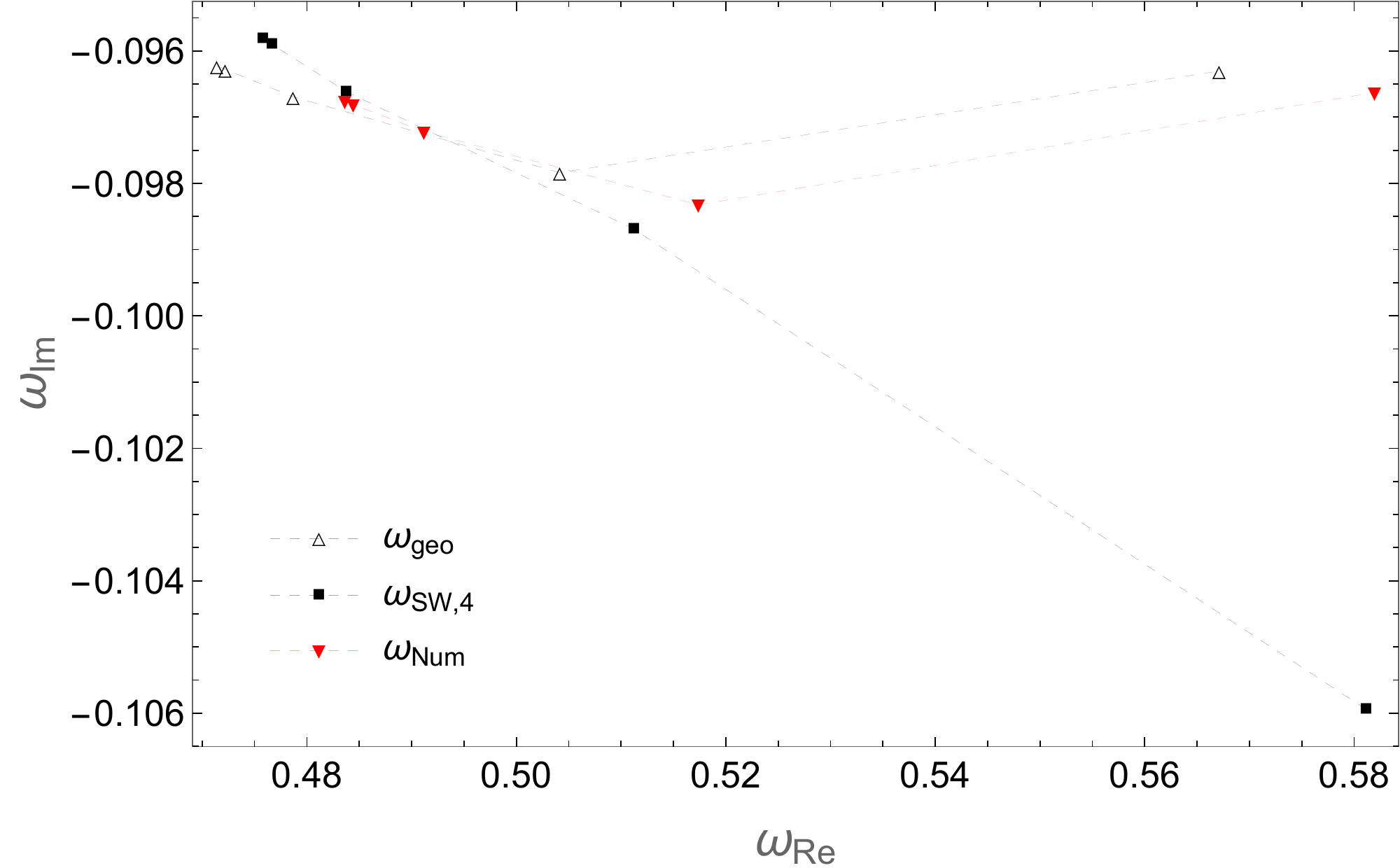}
\,
\includegraphics[scale=0.5]{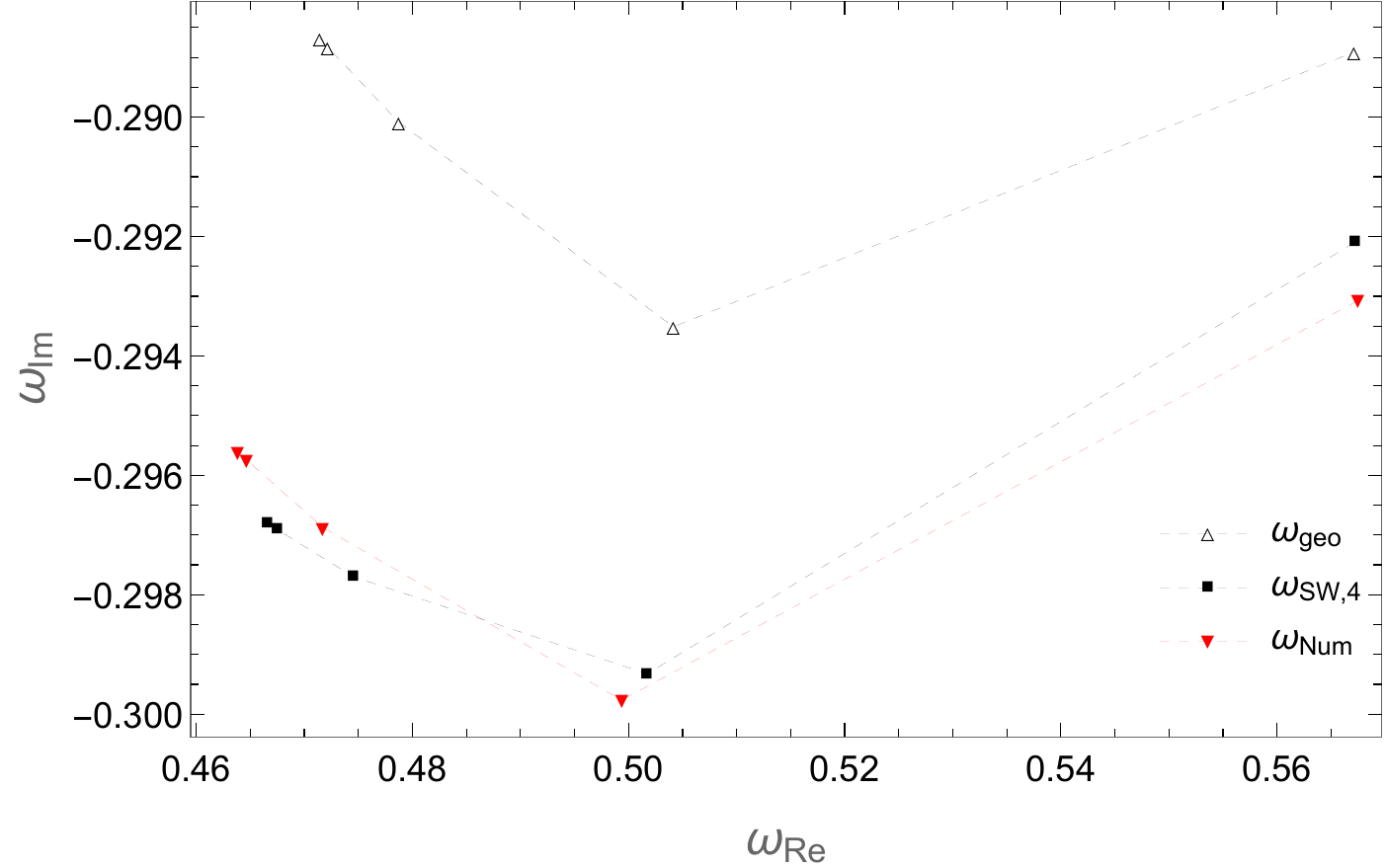}
\caption{QNMs of a RN BH for $n_\text{Num}=1$ (left) and $n_\text{Num}=2$ (right), $\ell=m_\phi=2$, with $a = 0$, $\mathcal{M} = 1$. $\mathcal{Q}$ varies between $0$ and $0.9$.}
\label{figplotsRN}
\end{figure}

\begin{figure}[t]
\center
\hspace{-1.5cm}
\includegraphics[scale=0.353]{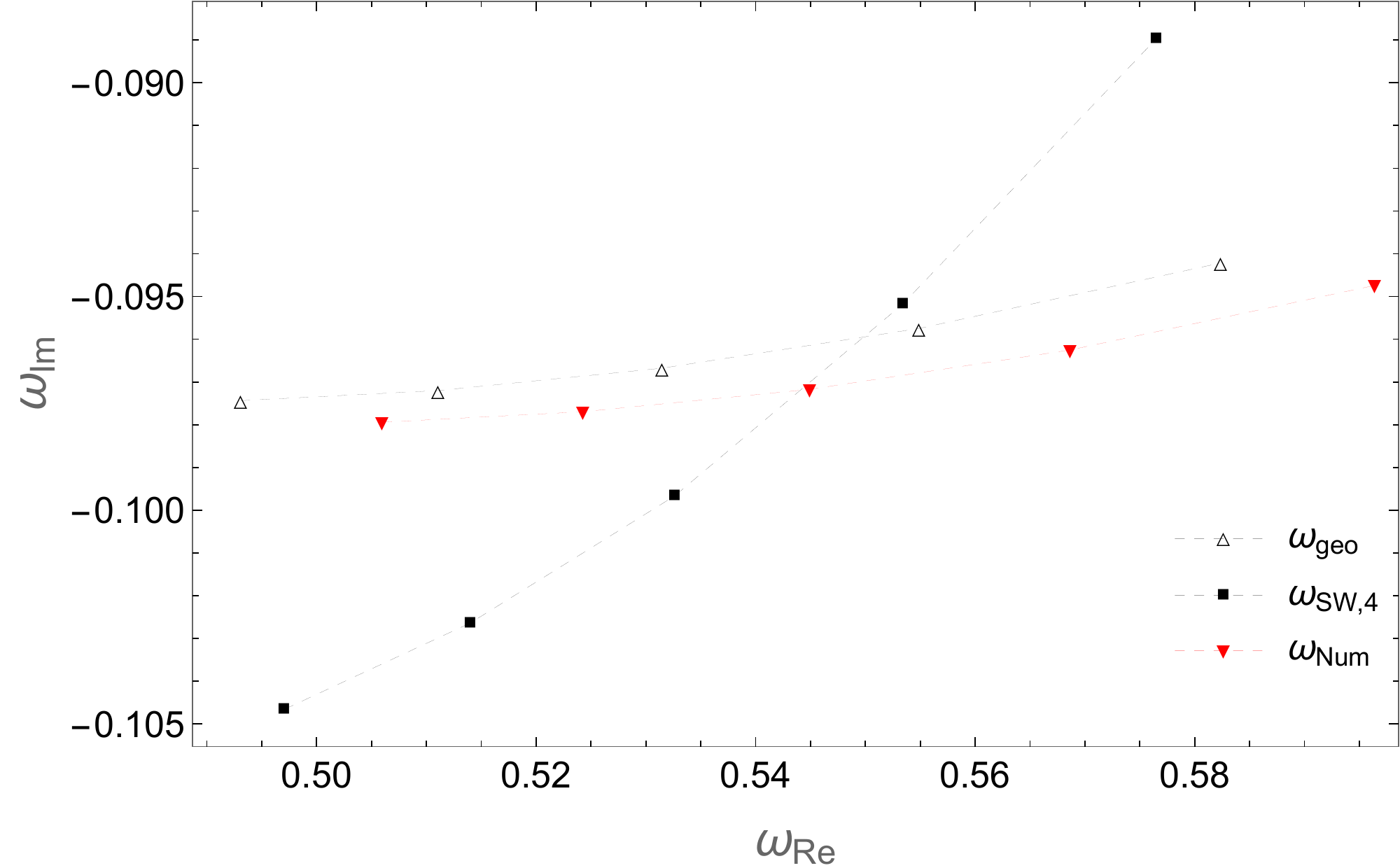}
\,
\includegraphics[scale=0.5]{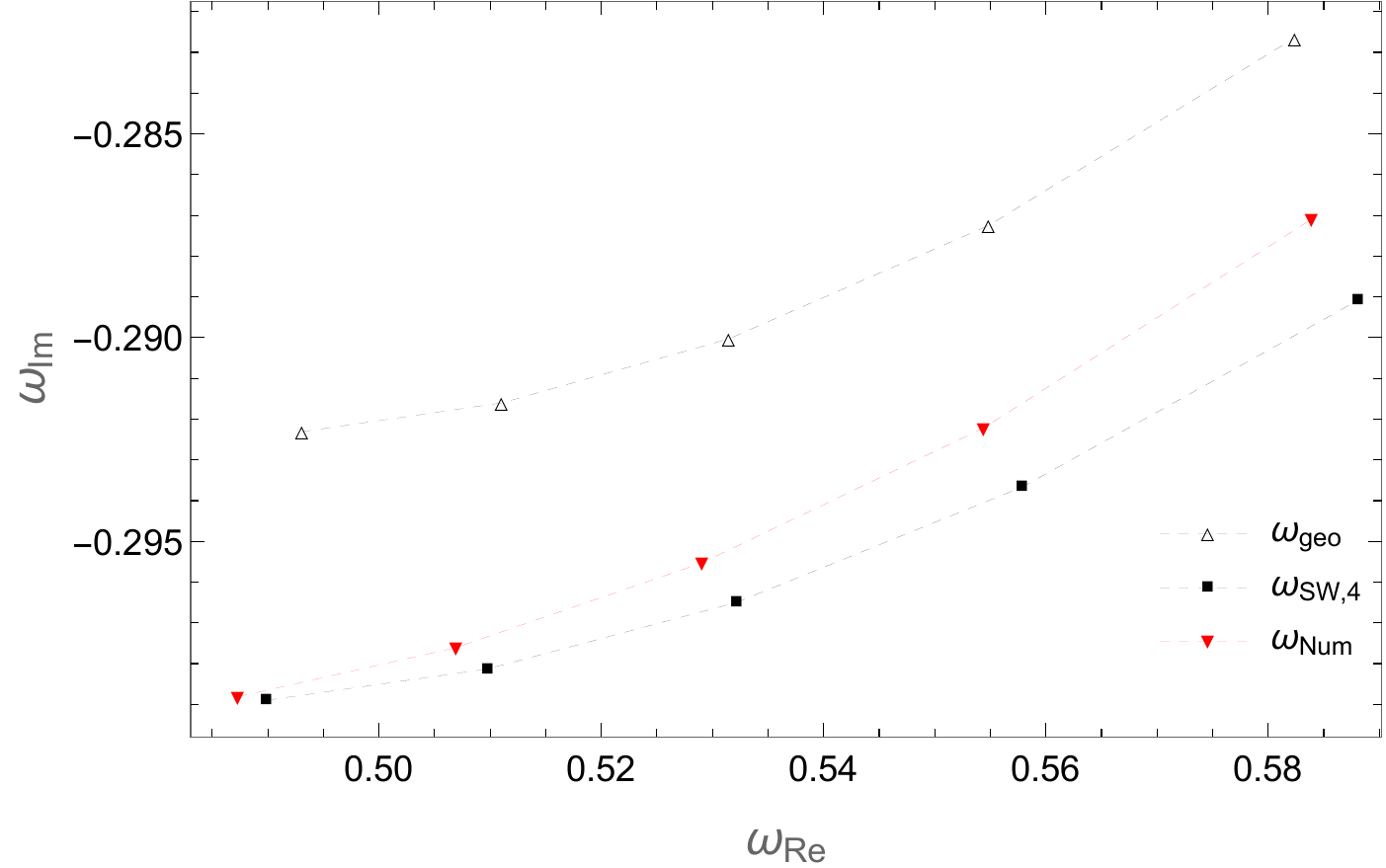}
\caption{QNMs of a KN BH for $n_\text{Num}=1$ (left) and $n_\text{Num}=2$ (right), $\ell=m_\phi=2$, with $\mathcal{Q} = 0.5$, $\mathcal{M} = 1$, $a_{_\mathcal{J}}$ varies between $0$ and $0.4$.}
\label{figplotsKNQ05}
\end{figure}

 Let us consider now the exact  SW quantization.  
 The radial and angular equations can be mapped to a $SU(2)$ theory with $N_f = (1,2)$ flavours.  We start from the expression (\ref{NS_prepotential_Nf_12}) for the NS prepotential up to four-instantons and invert the Matone relation (\ref{matone}) to compute $a$ as a function of $u$. The vanishing cycles for the radial and angular equations being $a_D$ and $a^\chi$, respectively, we impose the quantization conditions
\begin{equation}
\label{KN_cycles}
a_D = -\frac{1}{2\pi \rm{i}}\frac{\partial{\cal F}}{\partial a} = \hbar n_r
\quad , \quad
a^\chi -m_2^\chi = \hbar\left(  n_\chi + \frac{1}{2}\right)
\end{equation}
The two equations can be solved numerically for $\omega$ and $A$. We use Mathematica software, and to find the roots we look for solutions
around $\omega_{\rm geo}$  and $A_{\rm geo}$. Comparing against the WKB/geodetic results we find
\begin{equation}
n_r =-n_{\rm Num}= -(n_{r, geo} + 1)
\quad , \quad
n_\chi= \ell-|m| 
\end{equation}
where $n_{r, geo}$ starts from zero. Finally we compare the results with those provided by a continuous fraction numerical method \`a la Leaver.  The results of the three methods, along the relative error w.r.t. the numerical results for the Real and Imaginary part of the  frequencies, are shown in the following tables 
\begin{equation}
\begin{array}{||c||c|c||c|c||}
\hline
 n_\text{Num} = 1 & \omega  & A_{\text{lm}} & \text{Error}_{\%}(\omega _{\text{Re}} ) & E_{\%}(\omega _{\text{Im}} ) \\
\hline
 \text{Geo} & 0.555-0.0957 i & 5.99+0.0024 i & 2.42 & 0.524 \\
\hline
 \text{SW}_2 & 0.589-0.105 i & 5.99.+0.000651 i & 3.54 & 8.8 \\
\hline
 \text{SW}_4 & 0.557-0.0927 i & 5.99+0.00135 i & 2.07 & 3.68 \\
\hline
 \text{Num} & 0.569-0.0963 i & 5.99+0.00141 i & \text{} & \text{} \\
\hline
\end{array}
\end{equation}
and then for $n_r = 2$
\begin{equation}
\begin{array}{||c||c|c||c|c||}
\hline
 n_\text{Num} = 2 & \omega  & A_{\text{lm}} & \text{Error}_{\%}(\omega _{\text{Re}} ) & E_{\%}(\omega _{\text{Im}} ) \\
\hline
 \text{Geo} & 0.555-0.287 i & 5.99+0.0072 i & 0.0917 & 1.7 \\
\hline
 \text{SW}_2 & 0.545-0.3 i & 5.99+0.00212 i & 1.66 & 2.55 \\
\hline
 \text{SW}_4 & 0.558-0.294 i & 5.99+0.00408 i & 0.634 & 0.495 \\
\hline
 \text{Num} & 0.554-0.292 i & 5.99+0.00417 i & \text{} & \text{} \\
\hline
\end{array}
\end{equation}
In figures \ref{figplotsRN}, \ref{figplotsKNQ05} and \ref{figplotsKNExt} we display the results for the QNMs for various choices of the charge and angular momentum obtained using each of the three methods described in this paper. We set $\ell = m_\phi =2$, the mass $\cal M$ to one and find an amazing agreement between the three methods, even for low values of the energies and angular momenta, wherein the geodetic motion approximation is not expected to work. Tables collecting the data in the plots are presented in the appendix for the convenience of the reader. 

\subsection{Analytic solutions}
In this subsection we show two special cases of KN solutions related to perturbative gauge theories
\begin{itemize}
\item{ {\it static}: $\omega=0$}
\item{ {\it super-radiant}: $\omega=m_\phi \Omega_\phi$} with $\Omega_\phi = {a_{_\mathcal{J}} \over \mathcal{M}^2 + a_{_\mathcal{J}}^2}$ the angular velocity at the horizon
\end{itemize}
where analytic solutions can be found.
\subsubsection{Static Kerr-Newman wave}
 
Following the recent analysis for Kerr BH \cite{BonTanzetc}, here we consider a static wave ($\omega=0$) in the KN BH that, according to the gauge/gravity dictionary (\ref{dicKN}), is mapped to a gauge theory with $q=0$. It is easy to see that mixing terms depend on $a_{_\mathcal{J}}$ only through the combination $a_{_\mathcal{J}} \omega$, thus for $\omega=0$ angular and radial equations completely decouple and the separation constant simply becomes $A=\ell(\ell+1)$. The radial equation reduces to
\begin{equation}
R^{''}(r) + \frac{a_{_\mathcal{J}}^2 m_\phi^2 -A \,(r-r_+)(r-r_-)  + \tfrac{1}{4} (r_+-r_-)^2}{(r-r_+)^2(r-r_-)^2} \, R(r) = 0
\end{equation}
where $\Delta_r = (r-r_+) (r-r_-)$. The gauge/gravity dictionary reads (see \ref{dicKN})
\begin{equation}
q=0 \quad, \quad
\frac{u}{\hbar^2}= (\ell+\tfrac{1}{2})^2 \quad, \quad
m_1 = m_3 = 0 \quad, \quad
\frac{m_2}{\hbar}= \frac{2 i m_\phi \,a_{_\mathcal{J}}  }{r_+-r_-}  
\end{equation}
The general solution can be written as
\begin{equation}
\frac{R(r)}{\sqrt{\Delta_r}} = c_1\,P_{\ell, \mu} \left(1-\frac{2 \left(r-r_-\right)}{r_+-r_-}\right)+c_2\,Q_{-\ell-1,\mu}\left(1-\frac{2 \left(r-r_-\right)}{r_+-r_-}\right)
\end{equation}
with $P_{\ell, \mu}$ and $Q_{\ell,\mu}$ associated Legendre functions defined in (\ref{associated_Legendre_functions}) and 
\begin{equation}
\mu= -\frac{m_2}{\hbar} =-\frac{2{\rm i} m_\phi a_{_{\cal J}}}{r_+-r_-}
\end{equation}
For large $r$ the two solutions above behave like $r^\ell$ and $r^{-\ell-1}$ respectively. The static Love number is defined as the ratio of the coefficients of these terms \cite{Binnington:2009bb}. For a BH, one has to impose in-going boundary conditions at $r=r_+$. In this limit the contribution of $Q_{-1 -\ell,\mu}$ diverges so regularity requires $c_2=0$, leading to a zero static Love number as expected \cite{Binnington:2009bb,Damour:2009vw,LeTiec:2020bos}.
\subsubsection{Near super-radiant modes}
\label{NSRMs}

\begin{figure}[t]
\center
\hspace{-1.5cm}
\includegraphics[scale=0.353]{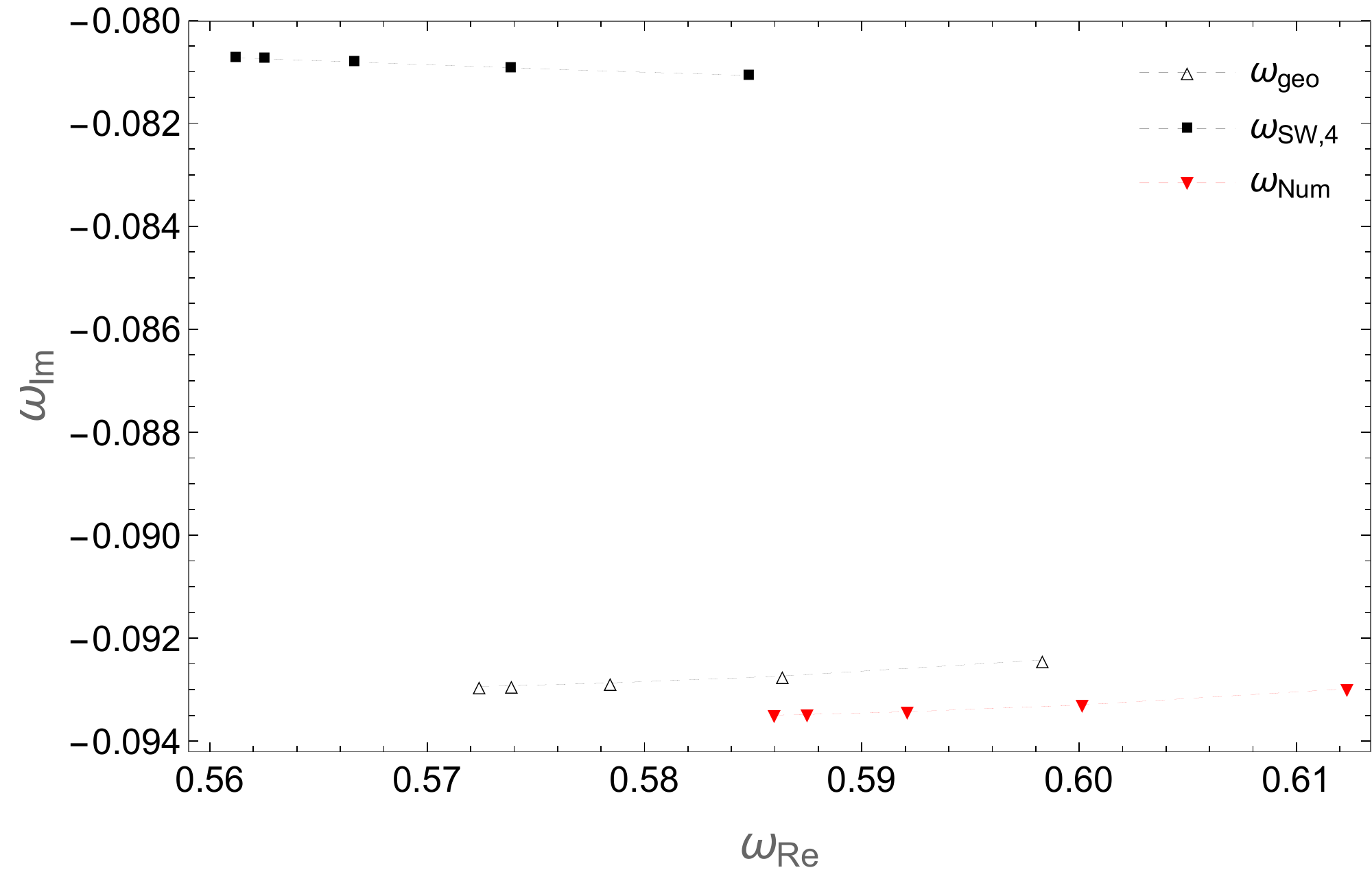}
\,
\includegraphics[scale=0.5]{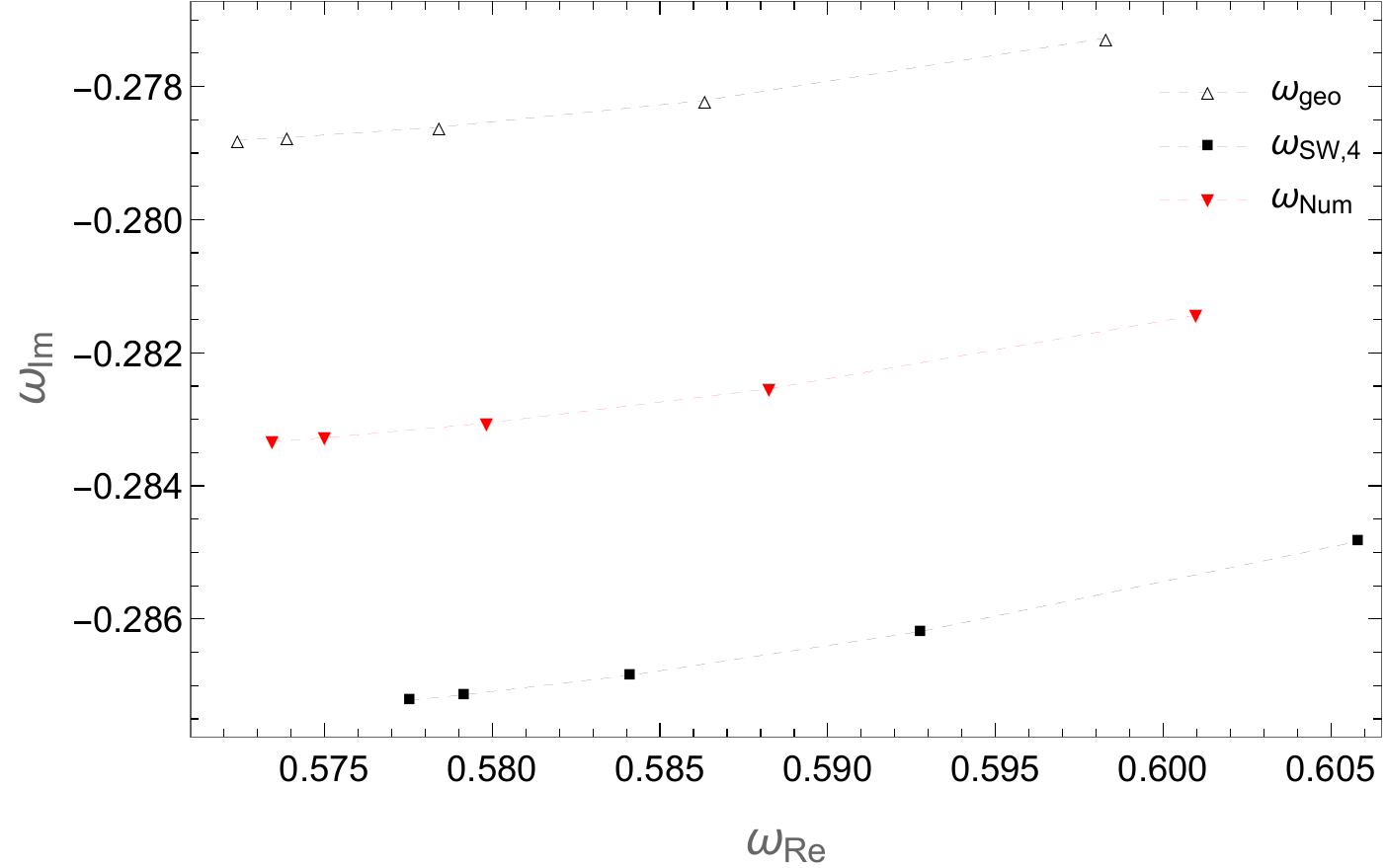}
\caption{QNMs of a KN BH for $n_\text{Num}=1$ (left) and $n_\text{Num}=2$ (right), $\ell=m_\phi=2$, with $a_{_\mathcal{J}} = 0.5$, $\mathcal{M} = 1$, $\mathcal{Q}$ varies between $0$ and $0.4$.}
\label{figplotsKNExt}
\end{figure}

In this subsection we consider near super-radiant modes, also known as zero-damping modes (ZDMs), which are close to the super-radiant threshold frequency $\omega_{_{SR}}$ \cite{Brito:2015oca} and the imaginary part of the frequency is almost vanishing. These modes are produced by near extremal BH mergers \cite{Starobinskil:1974nkd,Teukolsky:1974yv,Yang:2013uba}.

As a concrete example let's consider again the KN BH, whose QNM-SW dictionary is given in (\ref{dicKN}). 
Near super-radiant modes are defined by taking
\begin{equation}
\omega =  \omega_{_{SR}} + \nu {\delta}
\label{omegaNSR}
\end{equation}
with $\delta = r_+ - r_-$ the distance between the inner and outer horizon taken to be very small and
\begin{equation}
\omega_{_{SR}} = \frac{m_\phi\, a_{_\mathcal{J}}}{r_+^2+a_{_\mathcal{J}}^2} = m_\phi \Omega_\phi
\end{equation}
the super-radiant frequency. 

In order to compute $\nu$ one has to construct the QNM solutions in the near extremal limit. Far from the horizon $r \gg  r_+\gg \delta$ the radial equation reduces to
\begin{equation}
R{''}(r) + \left[\omega_{_{SR}}^2 + \frac{4 \omega_{_{SR}}^2 r_+}{r-r_+} + \frac{( 6 r_+^2+a_{_\mathcal{J}}^2)\omega_{_{SR}}^2 -A}{(r-r_+)^2}\right] R(r) = 0\,.
\end{equation}
The solutions to this equation are confluent hypergeometric functions. Requiring the absence of incoming waves one finds
\begin{equation}
\label{kerr_newman_extremal_wavefunction}
R(r) = c_\infty e^{i \omega_{_{SR}} (r-r_+)} (r-r_+)^{\frac{1}{2}+\alpha}  U\left(\tilde{A}; \tilde{B}; z_r\right)  
\end{equation}
where $c_\infty$ is a constant and $U$ is the Tricomi confluent hypergeometric function\footnote{Related to the Kummer confluent hypergeometric functions by
\begin{equation}
U(\tilde{A}; \tilde{B}; z) = \frac{\Gamma(1-\tilde{B})}{\Gamma(\tilde{A}-\tilde{B}+1)} {}_1F_1(\tilde{A}; \tilde{B}; z) + \frac{\Gamma(\tilde{B}-1)z^{1-\tilde{B}}}{\Gamma(\tilde{A})} {}_1F_1(\tilde{A}-\tilde{B}+1;2- \tilde{B}; z)
\end{equation}}, while
\begin{equation}
\tilde{A} = \frac{1}{2} +\alpha -2 i \omega_{_{SR}} r_+\,,
\qquad
\tilde{B} = 1+2\alpha\,,
\qquad
z_r = -2 i \omega_{_{SR}} (r-r_+)
\end{equation}
with  
\begin{equation}
\alpha = \sqrt{A + \frac{1}{4} - (a_{_\mathcal{J}}^2+6r_+^2) \,\omega_{_{SR}}^2}
\end{equation}
On the other hand, the radial equation in the near horizon limit can be 
approximated by taking $r = r_+ + \tau \delta$ and sending $\delta \to 0$.  In the variable $\tau$ one finds
\begin{equation}
\begin{aligned}
R{''}(\tau) + Q(\tau) R(\tau) = 0\,,
\end{aligned}
\end{equation}
with
\begin{equation}
Q(\tau) = \frac{\left( \nu(r_+^2{+}a_{_\mathcal{J}}^2)+2 r_+ \omega_{_{SR}} \tau \right)^2+\tfrac{1}{4}-\tau (\tau +1) (\alpha^2+4 r_+^2 \omega_{_{SR}}^2-\tfrac{1}{4})}{\tau ^2 (\tau +1)^2}
\end{equation}
The solution can be written in terms of ordinary hypergeometric functions. Imposing in-going boundary conditions at the horizon one finds
\begin{equation}
R(\tau) = c_H
\tau^{\frac{1}{2}-\frac{i a_{_\mathcal{J}} \nu }{\Omega _{\phi }}}
(1+\tau)^{\frac{1}{2}-\frac{i a_{_\mathcal{J}} \nu }{\Omega _{\phi }}+2 i r_+\omega_{_{SR}}}
{}_2 F_1 \left(\bar{A},\bar{B};\bar{C};-\tau\right)\label{r2}
\end{equation}
where $c_H$ is a constant and
\begin{equation}
\bar{A} =\frac{1}{2} -\alpha -\frac{2 {\rm i} a_{_\mathcal{J}} \nu }{\Omega _{\phi }}+2 {\rm i} r_+ \omega_{_{SR}}\,,
\qquad
\bar{B} =\frac{1}{2}+\alpha  -\frac{2 {\rm i} a_{_\mathcal{J}} \nu }{\Omega _{\phi }}+2 {\rm i}r_+ \omega_{_{SR}}\,,
\qquad
\bar{C} = 1-\frac{2 {\rm i} a_{_\mathcal{J}} \nu }{\Omega _{\phi }}
\end{equation}
By expanding (\ref{kerr_newman_extremal_wavefunction}) near the horizon $r\approx  r_+$ one finds  
\begin{equation}
R(r) \sim (r-r_+)^{\frac{1}{2}-\alpha } \left[1+(-2{\rm i}\omega_{_{SR}} )^{-2 \alpha} (r-r_+)^{2 \alpha }\frac{\Gamma (2 \alpha ) \Gamma (\bar{C}-\bar{B})}{\Gamma (-2 \alpha ) \Gamma (\bar{C}-\bar{A})} +\ldots \right] \label{eq1}
\end{equation}
while  far away from the horizon $r\gg r_+$, (\ref{r2}) reduces to
\begin{equation}
R(r) \sim (r-r_+)^{\frac{1}{2}-\alpha } \left[1+ \delta ^{2 \alpha }(r-r_+)^{2 \alpha }\frac{\Gamma (-2 \alpha )\Gamma (\bar{B}) \Gamma (\bar{C}-\bar{A})}{\Gamma (2 \alpha )\Gamma (\bar{A}) \Gamma (\bar{C}-\bar{B})}+\ldots \right]\label{eq2}
\end{equation}
by matching  (\ref{eq1}) and (\ref{eq2}) one finds
\begin{equation}
\label{near_super_radiant_quantization}
\left(-2 {\rm i} \omega_{_{SR}} \, \delta \right)^{-2\alpha}\frac{\Gamma (2 \alpha )^2 \Gamma (\bar{A}) \Gamma (\bar{C}-\bar{B})^2}{\Gamma (-2 \alpha )^2 \Gamma (\bar{B}) \Gamma (\bar{C}-\bar{A})^2} = 1
\end{equation}
Since $\delta \sim 0$ and ${\rm Re}\,\alpha > 0$,  the factor $\delta^{-\alpha}$ in the left hand side diverges, so it has to be compensated by a pole of  $\Gamma(\bar{B})$ in the denominator i.e.
\be
\bar{B} = - n + (-2 {\rm i}  \, \omega_{_{SR}}\delta )^{2\alpha} \eta
\ee
 with 
\begin{equation}
\eta = \frac{(-1)^n}{n!}\frac{\Gamma (-2 \alpha )^2 \Gamma \left(\bar{C}-\bar{A}\right)^2}{\Gamma (2 \alpha )^2 \Gamma \left(\bar{A}\right) \Gamma \left(\bar{C}-\bar{B}\right)^2}
\end{equation}
leading to \cite{Hod:2008se, Hod:2012bw} 
\bea
\omega &=&   \omega_{_{SR}} +  {\delta} \left[ { \Omega_\phi \, r_+ \, \omega_{_{SR}} \over  a_{_\mathcal{J}} }  -{\rm i} (n+\ft12+\alpha){\Omega_\phi\over 2 a_{_\mathcal{J}} }     \right] +\ldots \nn\\
&=& \Omega_\phi m_\phi (1 + 4\pi\,T_\text{BH}\,r_+ ) - 2\pi {\rm i}\, T_\text{BH}\,(\alpha   + n + \tfrac{1}{2}) + O\left(T_{BH}^{1+2\alpha} \right)
\eea
 This is consistent with the WKB analysis which states that the imaginary part of the quasi-normal frequencies is given by the Lyapunov exponent (\ref{QNMWKB}) that for near extremal rotating BHs the photon-sphere corresponding to co-rotating impinging photons coalesces with the horizon and $\lambda$ is proportional to the BH temperature \cite{Bianchi:2020des,Maldacena:2015waa}.

The resulting near super-radiant frequencies can be easily re-derived from exact SW quantization. In the limit $\delta \to 0$, the gauge coupling vanishes and the mass $m_2$ diverges unless we choose $\omega$ as in \eqref{omegaNSR}. In this particular limit the radial dictionary \eqref{dicKN} reduces to
\begin{equation}
\begin{aligned}
\frac{q}{\hbar} = 2 i \omega_{_{SR}} \delta \,, \qquad
\frac{u}{\hbar^2}= \alpha^2 \,,\qquad
\frac{m_1}{\hbar}=\frac{m_3}{\hbar}= -2 i \omega_{_{SR}} r_+\,,\qquad
\frac{m_2}{\hbar}=  2 i[r_+  \omega_{_{SR}}{-}\nu(a_{_\mathcal{J}}^2{+}r_+^2)]
\end{aligned}
\end{equation}
where we kept the leading terms in $\delta$.
For $q$ small, instanton contributions can be discarded but the 1-loop term must be kept and the quantization of $a_D$ reduces to
\begin{equation}
\label{near_super_radiant_SW_quantization}
\exp\left(-\frac{2\pi{\rm i }\,a_D}{\hbar}\right) = \left( - { q \over  \hbar}\right)^{\tfrac{2\sqrt{u}}{\hbar}}
\frac{\Gamma^2 ( 1+\tfrac{2\sqrt{u}}{\hbar} )}{\Gamma^2 ( 1-\tfrac{2\sqrt{u}}{\hbar} )}
\prod_{i=1}^{3}
{\Gamma\left({1\over 2}{+} {m_i- \sqrt{u}\over \hbar}  \right)  \over \Gamma\left({1\over 2}{+} {m_i+\sqrt{u}\over \hbar}  \right)}
=1
\end{equation}
which is precisely equation (\ref{near_super_radiant_quantization}) written in terms of the gauge variables. 

\section{Other examples}
\label{other_geometries}
Gauge/ gravity dictionaries similar to the one we have found for KN (AdS) can be found for other asymptotically flat or AdS gravity solution that admit a photon-sphere. We will consider in turn: D3-branes, BPS BHs from four intersecting stacks of D3-branes \cite{Cvetic:1995uj}, five-dimensional (asymptotically flat) CCLP BHs \cite{CCLP1,CCLP2}, including their extremal limits \cite{BMPV}, D1D5 circular fuzzballs \cite{Lunin:2001fv}, JMaRT smooth horizonless solutions \cite{JMaRT}, including their BPS limit, known as GMS solutions \cite{GMS1,GMS2}. The derivation of their QNMs following the steps described for KN (AdS) is straightforward but laborious and is beyond the scope of the present investigation.  

\subsection{D3-branes}
D3-branes arise as solutions in type IIB supergravity and are a very interesting example since they are dual to pure $SU(2)$ SYM with $N_f = (0,0)$. The dilaton-axion field of type IIB is decoupled from the metric and its equation of motion corresponds to the differential equation for a scalar perturbation on the D3 background, it can be mapped to the Mathieu equation by performing the change of variables $r = L\, e^{{\rm i} z}$ \cite{Cvetic:2000xz,Gubser:1998iu}. The spectrum of QNMs for this system has been studied first in \cite{Kurita:2002he} for small $\omega$ and then in general using SW curves in \cite{Bianchi:2021xpr}. The metric reads
\begin{equation}
\label{D3metric}
ds^2 = H(r)^{-{1\over 2}}  (-dt^2+d{\bf x}^2)  + H(r)^{{1\over 2}}   (dr^2 + r^2 d\Omega_5^2)
\end{equation} 
the radial equation can be put in canonical form with
\begin{equation}
\label{qhbar}
Q(r)= {  4  \omega^2( r^4+L^4) -r^2 (4 \ell(\ell+4) +15) \over 4 r^4}\
\end{equation}
and the gauge/gravity dictionary reads
\begin{equation}
\frac{q}{\hbar^4}  ={\omega^4\, L^4 \over  16}\,,
\qquad
{u \over \hbar^2}  = \left(\frac{\ell}{2}+1\right)^2\,;
\qquad
\hbar^2 y = {4 r^2 \over \omega^2 L^4 }
\label{D3_dic}
\end{equation}
We refer the reader to \cite{Bianchi:2021xpr} for the detailed study of QNMs of D3-branes, including a successful comparison of the three methods (WKB/geodesics, quantum SW curve, Leaver's continuous fraction) to compute them.
\subsection{Intersecting D3 branes BHs in four dimensions}
Other examples of 4-dimensional (BPS) BHs can be found in type IIB supergravity considering the intersection of four stacks of D3-branes. These geometries possess four different charges $\mathcal{Q}_i$ which, if taken equal, lead to an extremal Reissner-Nordstr{\"o}m BH. The line element reads \cite{Cvetic:1995uj}
\begin{equation}
ds^2= - f(r)\,dt^2 + f(r)^{-1} \left[dr^2 + r^2\left(d\theta^2 + \sin^2\theta d\phi^2\right)\right]
\end{equation}
where
\begin{equation}
f(r) = \prod_{i=1}^4 \left(1+\frac{\mathcal{Q}_i}{r}\right)^{-\frac{1}{2}}
\end{equation}
Thanks to spherically symmetry the angular wave-equation can be solved in terms of spherical harmonics, while the radial equation in canonical form is defined by
\begin{equation}
Q_r = \frac{\omega ^2 \prod_{i=1}^4 \left(\mathcal{Q}_i+r\right) -\ell(\ell+1) r^2}{r^4}
\end{equation}
The wave equation can be mapped to the quantum SW curve for $SU(2)$ with $N_f = (1,1)$ and
\begin{equation}
\begin{aligned}
{q \over \hbar^2}&\text{ = }-4 \omega ^2 \sqrt{\Sigma_4}
\, , \quad
{u \over \hbar^2} \text{ = }\left(\ell+\frac{1}{2}\right)^2-\omega ^2 \left(\Sigma_2+2 \sqrt{\Sigma_4}\right)
\\
{m_1 \over \hbar} &\text{ = }  \frac{i \omega\Sigma_3}{2\sqrt{\Sigma_4}}
\, , \quad
{m_3 \over \hbar}\text{ = }  \frac{i \omega\Sigma_1}{2}
\, ; \quad
y \hbar= -\frac{i r}{2  \omega  \sqrt{\Sigma_4}}
\end{aligned}
\end{equation}
where
\begin{equation}
\Sigma_n = \sum_{i_1<\cdots <i_n}^4 \mathcal{Q}_{i_1}\cdots \mathcal{Q}_{i_n}
\end{equation}
The entropy of the system is proportional to $\sqrt{\Sigma_4}$. In the limit where $\Sigma_4 \to 0$ the gauge coupling $q$ vanishes and $m_1$ diverges, 
so that one fundamental decouples and the resulting theory is $N_f =(1,0)$.
\subsection{CCLP five-dimensional BHs}
 
CCLP metrics describe rotating solutions of Einstein-Maxwell theory in $d=5$, with mass ${\cal M}$, charge ${\cal Q}$, and angular momentum parameters $\ell_1$, $\ell_2$. The line element reads \cite{CCLP1,CCLP2}
\begin{equation}
\begin{aligned}
ds^2 &= -dt^2-\frac{2 \mathcal{Q} \,\omega _2 }{\Sigma }(dt-\omega _1)+\Delta _t(dt-\omega_1)^2  +
\Sigma  \left(d\theta^2+\frac{r^2dr^2}{\Delta _r}\right) +
\\
&
 + d\psi^2 \cos^2\theta \,(r^2+\ell _2^2 )+d\phi^2 \sin ^2\theta \,(r^2+\ell _1^2)
\end{aligned}
\end{equation}
with the one-forms $\omega_{1,2}$ given by
\begin{equation}
\begin{aligned}
\omega _1 &= \ell_2 \,\cos ^2 \theta\,d\psi + \ell_1 \,\sin ^2\theta\, d\phi  
\quad , \quad
\omega _2  = \ell_1 \,\cos ^2 \theta\,d\psi + \ell_2 \,\sin ^2\theta\, d\phi  
\end{aligned}
\end{equation}
and the functions $\Delta _r$, $\Delta _t$ and $\Sigma$ given by
\begin{equation}
\begin{aligned}
\Delta _r &= (r^2 - r_+^2)(r^2 - r_-^2)
\, , \quad
\Delta _t = \frac{2 \mathcal{M} \Sigma -\mathcal{Q}^2}{\Sigma ^2}
\, , \quad
\Sigma = r^2+\ell _1^2\cos ^2\theta+\ell _2^2\sin ^2\theta
\end{aligned}
\end{equation}
These geometries, much like KN BH's, possess two horizons located at
\begin{equation}
r_\pm^2 =\widehat{{\cal M}} \pm \sqrt{\widehat{{\cal M}}^2 - \widehat{{\cal Q}}^2}
\end{equation}
with
\begin{equation}
\widehat{{\cal M}} = {\cal M} - \frac{\ell_1^2+\ell_2^2}{2}
\quad , \quad
\widehat{{\cal Q}} = {\cal Q} + \ell_1 \ell_2
\end{equation}
The wave-equation for a scalar perturbation in the CCLP background can be separated into radial and angular equations. Introducing the variables $z=r^2$ and $\xi=\cos^2\theta$, these can be brought into canonical form with
\begin{equation}
\begin{aligned}
Q_{z} &= \frac{z \left(\mathcal{L}_{\mathcal{M}}^2-\mathcal{L}_{\mathcal{Q}}^2\right)+2 \mathcal{L}_{\mathcal{Q}} \left(\widehat{\mathcal{Q}}\,\mathcal{L}_{\mathcal{M}}+\widehat{\mathcal{M}}\,\mathcal{L}_{\mathcal{Q}}\right)-\Delta _z \left(K^2-\omega ^2 (z+2 \mathcal{M})+4\right)+\Delta _z'{}^2}{4 \Delta _z^2}
\\
Q_{\xi } &= \frac{(1-\xi) \xi  \left(K^2+\omega ^2 \xi \left(\ell _1^2-\ell _2^2\right)\right)+1-(1-\xi) m_{\psi }^2-\xi  m_{\phi }^2}{4 (1-\xi)^2 \xi ^2}
\end{aligned}
\end{equation}
where $\Delta_z = (z-z_+)(z-z_-)$, $K^2$ is the separation constant, $m_\phi$, $m_\psi$ are the projections of the total angular momentum along two orthogonal 2-planes, and 
\begin{equation}
\mathcal{L}_{\mathcal{M}} = \ell _1 m_{\phi }+\ell _2 m_{\psi }-2 \mathcal{M} \,\omega
\, , \quad
\mathcal{L}_{\mathcal{Q}}= \ell _1 m_{\psi }+\ell _2 m_{\phi }+\mathcal{Q} \,\omega
\end{equation}
Both equations can be mapped to $SU(2)$ gauge theory with $N_f = (0,2)$ flavours. The  gauge/gravity dictionary for the radial wave equation reads
\begin{equation}
{q\over \hbar^2} = - \frac{\omega ^2 }{4}({z}_+{-}{z}_-)
\, , \;\;
{u\over \hbar^2} = \frac{1{+}K^2{-}\omega ^2 (z_+{+}2 \mathcal{M})}{4}
\, , \;\;
{m_{1,2} \over \hbar} =  {-}\frac{i}{2}\frac{\mathcal{L}_{\mathcal{M}}{\mp}\mathcal{L}_{\mathcal{Q}}}{\sqrt{z_+}{\pm}\sqrt{z_-}} 
\, ; \;\;
y = {z{-}z_-\over z_- {-} z_+}
\end{equation}
The dictionary for the angular part reads
\begin{equation}
{q^\xi\over \hbar^2}  = \frac{ \omega ^2 }{4} \left(\ell _1^2-\ell _2^2\right)
\, , \quad
{u^\xi \over \hbar^2}  = \frac{1}{4} \left(1+K^2+\omega ^2\ell_1^2\right)
\, , \quad
{m^\xi_{1,2} \over \hbar} = \frac{m_\phi \pm m_\psi}{2} 
\, ; \quad
y^\xi = -\xi
\label{D1D5_ang}
\end{equation}
Similarly to KN BH's, imposing the extremality condition $z_-= z_+$ (or $\widehat{\cal M}^2 = \widehat{\cal Q}^2$) leads to the decoupling of the flavour associated to $m_2$. As $m_2\to \infty$ we keep $q_\text{ext} = - q\,m_2$ finite, after rescaling $y_\text{ext} = -y/m_2$ one is left with an $N_f = (0,1)$ theory with
\begin{equation}
\begin{aligned}
\frac{q_\text{ext}}{\hbar^3} &= - \frac{{\rm i}\,\omega ^2\sqrt{z_+}\left[(\ell _1+\ell _2)( m_{\phi }+ m_{\psi })-(2 \mathcal{M}-\mathcal{Q}) \,\omega \right]}{4}\,,
\qquad
{u\over \hbar^2}  = \frac{1{+}K^2{-}\omega ^2 (\widehat{{\cal M}}{+}2 \mathcal{M})}{4}
\\
{m_1 \over \hbar} &= \frac{(\ell_1-\ell_2)(m_\phi-m_\psi)-(2 \mathcal{M}+\mathcal{Q}) \omega}{4{\rm i}\sqrt{z_+}}\,;
\qquad
\frac{y_\text{ext}}{\hbar^2} = {z - z_+\over {\rm i}\sqrt{z_+}\left(\mathcal{L}_{\mathcal{M}} + \mathcal{L}_{\mathcal{Q}}\right)}
\label{D1D5_extr_radial}
\end{aligned}
\end{equation}
A particular case is the BMPV BH which is obtained by imposing also the BPS condition ${\cal M} = {\cal Q}$ (which implies $\ell_2 = -\ell_1$) \cite{BMPV}. The radial dictionary \eqref{D1D5_extr_radial} reduces to
\begin{equation}
\begin{aligned}
{q_\text{ext}\over \hbar^3} = \frac{i {\cal M}\sqrt{z_+}\omega^3}{4}
\, , \quad
{u\over \hbar^2}  &= \frac{1{+}K^2{-}\omega ^2 (3 \mathcal{M} - \ell_1^2)}{4}
\, , \quad
{m_{1} \over \hbar} =  \frac{i\left[\ell _1 (m_{\psi } - m_{\phi })+\tfrac{3 \omega  \mathcal{M}}{2}\right]}{2 \sqrt{z_+}}
\end{aligned}
\end{equation}
The dual gauge theory of the angular equation of a BMPV BH is still $N_f = (0,2)$, the dictionary follows from \eqref{D1D5_ang} and reads
\begin{equation}
{q^\xi\over \hbar^2}  = 0
\, , \quad
{u^\xi \over \hbar^2}  = \frac{1}{4} \left(1+K^2+\omega ^2\ell_1^2\right)
\, , \quad
{m^\xi_{1,2} \over \hbar} = \frac{m_\phi \pm m_\psi}{2} 
\, ; \quad
y^\xi = -\xi
\end{equation}
Notice that the angular equation relates to free $SU(2)$ SYM, therefore the exact solutions are known and $K^2=\ell(\ell+2)$.

At variant with KN BH's, there is another extremal limit that leads to pure $SU(2)$ SYM with no flavour obtained by imposing the vanishing of $z_+$, i. e. $2{\cal M}{\,=\,}\ell_1^2+\ell_2^2$ and ${\cal Q}{\,=\,}-\ell_1 \ell_2$. In this limit the mass $m_1$ diverges and while $\hat{q}_\text{ext} = -q_\text{ext}\,m_1$ is finite. The resulting geometry is associated to a gauge theory with $N_f=(0,0)$ and
\begin{equation}
\frac{\hat{q}_\text{ext}}{\hbar^4} = \frac{\omega ^2}{16}(\mathcal{L}_\mathcal{M}^2-\mathcal{L}_\mathcal{Q}^2)\,, \quad
{u\over \hbar^2}  = \frac{1{+}K^2{-}(\ell_1^2+\ell_2^2) \omega ^2}{4}\,;
\quad
\hat{y}_\text{ext}\,\hbar^2  = 
\frac{4z}{\mathcal{L}_\mathcal{M}^2-\mathcal{L}_\mathcal{Q}^2}
\label{CCLP_nf0_radial}
\end{equation}
As for the D3D3D3D3 BHs vanishing entropy corresponds to a theory with fewer flavours.
\subsection{D1D5 fuzzball} 
Next we consider a D1D5 circular fuzzball with radius $a_f$ and equal charges ${\cal Q}_1={\cal Q}_5=L^2$. The smooth horizonless metric is given by \cite{Lunin:2001fv}
\begin{equation}
\begin{aligned}
ds^2 &= H_f^{-1}\left[(d{v}+\omega_\psi\,d\psi)^2-(dt+\omega_\phi\,d\phi)^2\right]
+
\\
&+
H_f\left[d\phi^2 \sin ^2\theta (\rho ^2+a_f^2)+
\frac{\Sigma_f}{\rho ^2+a_f^2}\left[d\rho^2+(\rho ^2+a_f^2)d\theta^2 \right]+
\rho ^2 d\psi^2 \cos ^2\theta\right]
\end{aligned}
\end{equation}
with
\begin{equation}
\omega_\phi = \frac{L^2 a_f \sin ^2\theta}{\Sigma_f}
\, , \quad
\omega_\psi = \frac{L^2 a_f  \cos ^2\theta}{\Sigma_f}
\, , \quad
H_f = 1 + \frac{L^2}{\Sigma_f}
\, , \quad
\Sigma_f = \rho^2 + a_f^2 \cos^2\theta
\end{equation}
Setting
\begin{equation}
\Phi = e^{-i\omega t + i P_{v} v + i m_\phi \phi + i m_\psi \psi} R(\rho) S(\chi)
\end{equation}
the wave equation can be separated, brought into canonical form (\ref{secondD}) and matched to that of $SU(2)$ gauge theory with $N_f=(0,2)$ fundamentals. The $Q$-functions read
\begin{equation}
\begin{aligned}
Q_{\text{D1D5},\rho } &= \frac{\left(a_f^2-\rho ^2\right)^2+4 \left[\rho ^2 \mathcal{L}_{\phi }^2-\left(a_f^2+\rho ^2\right) \left(\mathcal{L}_{\psi }^2+\rho ^2\left(1+K^2-\left(2 L^2+\rho ^2\right) \tilde{\omega }^2\right)\right)\right]}{4 \rho ^2 \left(a_f^2+\rho ^2\right)^2}
\\
Q_{\text{D1D5},\chi } &= \frac{\left(\chi ^2+1\right)^2-4 \left[\chi ^2 m_{\phi }^2+\left(1-\chi ^2\right) \left(m_{\psi }^2-\chi ^2 \left(1+K^2+\tilde{\omega }^2 a_f^2 \chi ^2\right)\right)\right]}{4 \chi ^2 \left(1-\chi ^2\right)^2}
\end{aligned}
\end{equation}
where $\chi= \cos \theta$ and we defined
\begin{equation}
\mathcal{L}_{\phi } = a_f\,m_{\phi }-L^2 \omega
\quad , \quad
\mathcal{L}_{\psi } = a_f\,m_{\psi }-L^2 P_{v}
\quad , \quad
\tilde{\omega}^2 = \omega^2 - P_{v}^2
\end{equation}
The gauge/gravity dictionary for the radial equation reads
\begin{equation}
\begin{aligned}
{q\over \hbar^2} = \frac{a_f^2 \tilde{\omega}^2}{4}
\, , \quad
{u\over \hbar^2} = \frac{1+K^2+\tilde{\omega }^2 \left(a_f^2-2 L^2\right)}{4}
\, , \quad
{m_{1,2} \over \hbar} = \frac{{\cal L}_\phi \mp {\cal L}_\psi}{2a_f}
\, ; \quad
y = \frac{\rho^2}{a_f^2}
\end{aligned}
\end{equation}
while for the angular equation one finds 
\begin{equation}
\begin{aligned}
{q^\chi \over \hbar^2}= \frac{a_f^2 \tilde{\omega}^2}{4}
\quad , \quad
{u^\chi \over \hbar^2} = \frac{1+K^2+\tilde{\omega }^2 a_f^2}{4}
\quad , \quad
{m_{1,2}^\chi \over \hbar}= \frac{m_\phi \pm m_\psi}{2}
\quad ; \quad
y^\chi = -\chi^2
\end{aligned}
\end{equation}
In the BH limit $a_f = 0$ the gauge coupling goes to zero while both masses diverge ($q_\text{BH}= m_1 m_2 \,q$ is finite), the resulting theory is $N_f = (0,0)$ with radial dictionary
\begin{equation}
\begin{aligned}
\frac{q_\text{BH}}{\hbar^4}= \left(\frac{L \tilde{\omega}}{2}\right)^4
\quad , \quad
{u \over \hbar^2} = \frac{1+K^2-2\tilde{\omega }^2 L^2 }{4}
\quad ; \quad
\hbar^2\,y_\text{BH} = \frac{4\rho^2}{\tilde{\omega}^2 L^4}
\end{aligned}
\end{equation}
As for D3-branes the wave equation in this case can be solved exactly.
\subsection{JMaRT and GMS geometries} 
JMaRT solutions are (non-)BPS smooth horizonless geometries with three charges ${\cal Q}_1$, ${\cal Q}_5$ and ${\cal Q}_P$ and two angular momenta ${\cal J}_\phi$ and ${\cal J}_\psi$. 
The explicit form of metric and the other field profiles can be found in the original paper \cite{JMaRT}. The charges and the angular momenta are given by
\begin{equation}
\mathcal{M} = \sum_i \frac{c_i^2+s_i^2}{2}\,M \quad , \quad
\mathcal{Q}_i = M c_i s_i \quad , \quad
\mathcal{J}_\phi = - m \frac{ \mathcal{Q}_1 \mathcal{Q}_5 }{R_{v}} \quad , \quad
\mathcal{J}_\psi = n \frac{ \mathcal{Q}_1 \mathcal{Q}_5 }{R_{v}}
\end{equation}
with $i=1,5,P$ and  $c_i =\cosh\delta_i$ ,$s_i=\sinh\delta_i$ `boost' parameters satisfying $c_i^2-s_i^2=1$. The charges can be parametrized as
\begin{equation}
m-n = {j+ j^{-1} \over s + s^{-1}} \quad , \quad 
m+n = {j- j^{-1} \over s - s^{-1}} 
\end{equation}
with
\begin{equation}
j = \sqrt{a_2\over a_1}\le 1 \quad , \quad 
s = \sqrt{ s_1 s_5 s_P \over c_1 c_5 c_P}
\end{equation}
and
\begin{equation}
\mathcal{Q}_1 = {g_s\alpha'^3 \over V_4}n_1 \quad , \quad 
\mathcal{Q}_5 = g_s\alpha' n_5 \quad , \quad 
\mathcal{Q}_P = \frac{g_s^2\alpha'^4 n_P}{R_v^2 V_4}
\end{equation}
Note that the volume of the internal 4-torus is ${\cal V}(T^4) = (2\pi)^4 V_4$ while regularity requires $n_P = {n\, m \, n_1\, n_5}$\footnote{For simplicity we consider the non-orbifold case.}.

The scalar wave equation was separated in \cite{JMaRT} we will mostly follow their notations.
As we will show both radial and angular equations can be mapped to the differential equation associated to an $N_f=(0,2)$ theory.

Setting
\begin{equation}
\Phi(t,y,r,\theta,\psi,\phi) = e^{-i \omega t -i P_{v} {v}+i m_\psi \psi +i m_\phi \phi } H(r) \Theta (\theta)
\end{equation}
the radial and angular equations for $H(r)$ and $\Theta(\theta)$ separate. Using
\begin{equation}
z= \frac{r^2 - r_+^2 }{ r_+^2 - r_-^2} \quad, \quad
\end{equation}
where
\begin{equation}
r_\pm^2 = \frac{M-a_1^2-a_2^2}{2}\pm \frac{1}{2} \sqrt{(M-a_1^2-a_2^2)^2-4 a_1^2 a_2^2}
\end{equation}
The radial equation reads
\begin{equation}
4 \, \frac{d}{dz} \left[z(z+1) \frac{dH}{dz} \right] +
\left[ 1-\nu^2 + \kappa^2 z + \frac{\alpha^2}{z+1} - \frac{\beta^2}{z}\right] H(z) = 0
\end{equation}
Performing the redefinition 
\begin{equation}
H(z) = {F(z)\over \sqrt{z(z+1)}}
\end{equation}
we get the radial $Q$-function in canonical form
\begin{equation}
Q_\text{JMaRT}(z) =
\frac{\kappa^2 z^3+z^2 (\kappa^2-\nu ^2+1)+z (\alpha ^2+\beta ^2-\nu ^2+1)+1+\beta^2}{4 z^2(z+1)^2}
\label{Q_JMaRT_rad}
\end{equation}
where
\begin{equation}
\begin{aligned}
&\kappa^2 = \tilde{\omega}^2 (r_+^2 - r_-^2)   \\
&\nu^2 = 1+{{K^2}} + \tilde{\omega}^2(r_+^2+Ms_1^2+Ms_5^2) + M (\omega c_P + P_{v} s_P )\\
&\alpha = \omega R_{v} \sigma + P_{v} R_{v} \tau - m_\phi n - m_\psi m \\
&\beta =  P_{v} R_{v} \tau + m_\phi m - m_\psi n 
\label{JMaRT_pars}
\end{aligned}
\end{equation}
with $\tilde{\omega}^2=\omega^2 -P_{v}^2$ and
\begin{equation}
\sigma = {c_1^2c_5^2c_P^2 - s_1^2s_5^2s_P^2 \over c_1s_1c_5s_5} >0 \quad , \quad 
\tau = {(c_1^2c_5^2- s_1^2s_5^2)c_Ps_P \over c_1s_1c_5s_5} >0
\end{equation}
From eq. \eqref{Q_JMaRT_rad} we can find the dictionary with an $N_f=(0,2)$ theory \textit{viz.}
\begin{equation}
{q\over \hbar^2}= \frac{\kappa^2}{4} \quad, \quad
{u\over \hbar^2}= \frac{\nu^2-\kappa^2}{4} \quad, \quad
{m_1\over \hbar}= \frac{\alpha+ i \beta}{2} \quad, \quad
{m_2\over \hbar}= \frac{\alpha- i \beta}{2} \quad; \quad 
y_{SW}= z 
\end{equation}
The angular equation reads
\begin{equation}
\frac{1}{\sin 2\theta} \frac{d}{d\theta} \left(\sin 2\theta \frac{d\Theta}{d\theta}\right) +
\left[ K^2+ \tilde{\omega}^2 (a_2^2 \cos^2\theta + a_1^2\sin^2\theta) - {m_\psi^2 \over \cos^2\theta}
- {m_\phi^2 \over \cos^2\theta}\right]\Theta = 0
\end{equation}
that can be put in canonical form by setting $\xi= \cos^2\theta$ and
\begin{equation}
\Theta = \frac{S(\xi)}{\sqrt{\xi(1-\xi)}}
\end{equation}
the $Q$-function reads
\begin{equation}
Q_\text{JMaRT}^{\theta} = {1+K^2\xi(1-\xi) - m^2_\phi \xi - m^2_\psi(1-\xi) +\tilde\omega^2 \xi(1-\xi)
[a_2^2\xi + a_1^2(1-\xi)] \over 4\xi^2(1-\xi)^2}
\end{equation}
Again, the above potential matches the one of the $N_f=(0,2)$ theory with the following dictionary
\begin{equation}
{q^\xi \over \hbar^2}= -\frac{(a_1^2{-}a_2^2)\tilde{\omega}^2}{4} \quad, \quad
{u^\xi \over \hbar^2}= \frac{1{+}K^2}{4}+\frac{a_2^2 \tilde{\omega}^2}{4} \quad, \quad
{m_{1,2}^\xi \over \hbar}= \frac{m_\phi{\pm} m_\psi}{2} \quad; \quad
y_{SW}^\xi= -\xi
\end{equation}

The GMS geometry \cite{GMS1,GMS2} can be obtained from the JMaRT geometry setting $m=n+1$.

It is worth mentioning that the analysis performed in \cite{Cardoso:2005gj,Chakrabarty:2015foa,Eperon:2016cdd,Chakrabarty:2019ujg} on the spectrum of QNM in either the eikonal limit or when $\tilde{\omega} = 0$ corresponds, in the gauge theory, to $q$ going to infinity or zero, respectively. As for the extremal KN case discussed in section \ref{NSRMs} an analytic derivation of the QNMs can be provided, which completely agrees with the quantization of the $a_D$ cycle on the gauge theory side, by solving the wave equation in different regions and imposing matching conditions and the correct boundary conditions.
\section{Conclusions and outlook}
\label{conclusions}

Let us conclude by summarising the results reached by the present investigation and speculating about the origin of the QNM-SW gauge/gravity correspondence.

After reviewing the three available approaches \textit{i.e.} WKB/geodesics, numerical methods \`a la Leaver and quantum SW curves, we have exploited them to compute the QNMs of massless scalar perturbations of KN BHs with arbitrary charge $\mathcal{Q}$ and angular momentum $\mathcal{J}= a_{_\mathcal{J}}\mathcal{M}$. The values are displayed in the plots in figures \ref{figplotsRN}, \ref{figplotsKNQ05}, \ref{figplotsKNExt} and in the tables in Appendix \ref{appendixTables}. The agreement with the numerical results is remarkable. Even more remarkable is the agreement with the geodesics/WKB approach for low values of the charges where the semi-classical approximation is not fully justified.

Moreover we have illustrated the procedure for various toy models: the 'inverted hydrogen atom', 'spherical harmonics' , static KN waves, and near super-radiant modes of quasi-extremal KN BHs. These cases are related to free gauge theories and admit an analytic solution allowing us to streamline the prescription to identify the cycle $\gamma$ whose period has to be quantized with the one shrinking to zero size in the classical limit $\hbar \rightarrow 0$.

We have also shown how different gauge theories, \textit{e.g.} with the same $SU(2)$ group but different number of flavours, may capture the same QNMs. We exploited this to rewrite differential equations with two irregular singular points (aka doubly-confluent Heun equation DCHE) as differential equations with two regular singular points and one irregular singular point (aka confluent Heun equation CHE). The prototypical case with $N_f=4$ that capture the QNMs of KN-AdS BH is governed by the 'standard' Heun equation with four regular singular points.

Finally we have established the detailed gauge / gravity dictionaries  for various classes of BHs, branes and fuzz balls in diverse dimensions. Quite surprisingly the elliptic geometry associated to an $SU(2)$ gauge group combined with various choices of flavours is sufficient to encompass systems ranging from (intersecting) D3-branes and their BPS bound-states, CCLP BHs in $D=5$ and their extremal limits, D1-D5 circular fuzz balls, JMaRT smooth horizonless geometries and their BPS limit aka GMS. The computation of their QNMs following the steps described earlier on looks feasible though somewhat laborious. Each case deserves a dedicated analysis. 
 
    In this work we exploit the mathematical equivalence among \textit{\it a priori} completely unrelated systems, a four-dimensional gauge theory and a BH (or a brane) solution. M-theory may provide a hint on the physical and geometrical origin of this correspondence. 
   Gauge theories with  $\mathcal{N}=2$ supersymmetry on a NS background can be  realized by wrapping M5-branes on Riemann surfaces that coincide with the SW curves themselves embedded on a non-commutative space  \cite{Fucito:2011pn}. It is tempting to speculate that the very same Riemann surface representing a homological 2-cycle in a local CY geometry, wrapped by a stack of M2-branes may provide the M-theory description of the corresponding BHs. The two systems are related by electromagnetic duality in eleven dimensions.  The near photon-sphere dynamics would be captured by some CFT that might well be the AGT dual of the $\mathcal{N}=2$ SYM theory, i.e. the reduction of the M5 brane theory along the four-dimensional gauge theory spacetime effectively compactified by the presence of the $\Omega$-background. For extremal Kerr and KN BHs a form of the holographic correspondence between near-horizon extremal Kerr (NHEK) and 2-d chiral CFT  \cite{Guica:2008mu,Compere:2012jk} has already been exploited in the study of (metric) perturbations in Kerr-AdS and its near-horizon geometry \cite{Dias:2012pp}. Further investigation is required to push these ideas onto firmer grounds. 

  We remark that the methods of the present investigation allow to compute not only the spectrum of QNMs but also other physical properties of these geometries, that can help to discriminate between BHs and smooth horizonless compact objects.

\section*{Acknowledgments}

We would like to thank A. Aldi, C. Argento, G. Bonelli,
V. Cardoso, G. Di Russo, D. Fioravanti, M. Firrotta,
F. Fucito, A. Grassi, T. Ikeda, C. Iossa, M. Mari\~no,  D. Panea Lichtig, P. Pani,
G. Raposo, R. Savelli, A. Tanzini and Y. Zenkevich for interesting
discussions and valuable suggestions.

\begin{appendix}

\section{The one-loop prepotential}

\label{appendixA}

The one-loop prepotential for $SU(2)$ gauge theory with $N_f=4$ fundamentals is given by
\bea
{\cal F}_{\rm 1{-}loop} (a) &=&   \epsilon_1\, \epsilon_2 
\log {  \prod_{i=1}^4 \Gamma_2\left(m_i -a +{\epsilon\over 2}\right) \Gamma_2\left(m_i +a+{\epsilon\over 2}\right) \over   \Gamma_2(2a+\epsilon)\Gamma_2(-2a+\epsilon)  }  \label{f1loop}
\eea      
with $\Gamma_2$ the Barnes double gamma function defined by the integral\footnote{In the following and in the main text we used the shorthand notation $\Gamma_2(x)$ when it is not necessary to specify its dependence on the $\epsilon_i$. The related function $\gamma_{\epsilon_1,\epsilon_2}(x)=\Gamma_2(x+\epsilon)$ is often used in the literature.}
\be
\log\Gamma_2(x|\epsilon_1,\epsilon_2)  = {d\over ds}\left(  {\Lambda^s\over \Gamma(s) } \int_0^\infty {dt\over t}
{t^s\, e^{-x t} \over
(1-e^{-\epsilon_1 t} ) (1-e^{-\epsilon_2 t} )  }\right)\Big|_{s=0}  \label{ge1e2}
\ee
and with $\epsilon=\epsilon_1+\epsilon_2$. In the limit $\epsilon_2 \to 0$ one finds
\bea
\lim_{\epsilon_2\to 0} \epsilon_2 \partial_x \log\Gamma_2(x|\epsilon_1,\epsilon_2) & =&  -{d\over ds}\left(  {\Lambda^s\over \Gamma(s) } \int_0^\infty {dt\over t}
{t^s\, e^{-x t} \over
1-e^{-\epsilon_1 t} }\right)\Big|_{s=0}  = -{d\over ds} \left[ \epsilon_1^{1-s} \Lambda^s \zeta_s\left({x\over \epsilon_1}\right) \right]_{s=0} \nn\\
&= & - \epsilon_1 \log\Gamma\left({x \over \epsilon_1} \right) - x \log\left({ \epsilon_1 \over \Lambda} \right) +{\epsilon_1\over 2} 
\log\left({2\pi \epsilon_1 \over \Lambda} \right)  \label{ge1e22}
\eea 
with $\zeta_s$ the Hurtwitz zeta function. 
Plugging this into (\ref{f1loop}) and setting $\epsilon_1= \hbar$ one finds the NS one-loop prepotential
   \bea
{\partial {\cal F}_{\rm 1{-}loop} (a) \over \partial a} &=&  2\hbar \log    { \Gamma\left(1+ {2a\over \hbar}  \right)  \over \Gamma\left(1- {2a\over \hbar}  \right) }  + \hbar \sum_{i=1}^{4}
 \log    { \Gamma\left({1\over 2}+ {m_i-a\over \hbar}  \right)  \over \Gamma\left({1\over 2}+ {m_i+a\over \hbar}  \right) }  
 \eea     
 
\section{\texorpdfstring{$U(y)$}{U(y)} vs \texorpdfstring{${\widetilde U}(x)$}{U(x)} at \texorpdfstring{$q=0$}{q=0}}
\label{appendixB}

In this appendix we show how the wave functions $U(y)$ and ${\widetilde U}(x)$ are related in the free gauge theory case where all the calculations can be performed analytically.

We first solve \eqref{difq0} at $q=0$
\begin{equation}
U''(y)+
\frac{y \hbar -m_1-m_2}{y (y+1) \hbar }\, U'(y)+
\frac{( m_1+\tfrac{\hbar}{2} ) ( m_2+\tfrac{\hbar}{2})-u y}{y^2 (y+1) \hbar^2}\,U(y) = 0
\label{diffeq_y_q0}
\end{equation}
The general solutions are hypergeometric functions
\begin{equation}
\begin{aligned}
U(y)=&\, d_1\, y^{\frac{1}{2}+\frac{m_1}{\hbar}} \,{}_2 F_1(\tfrac{1}{2}+\tfrac{m_1-\sqrt{u}}{\hbar},\tfrac{1}{2}+\tfrac{m_1+\sqrt{u}}{\hbar},1+\tfrac{m_1-m_2}{\hbar}|-y) 
\\
+ &\, d_2 \, y^{\frac{1}{2}+\frac{m_2}{\hbar}} \,{}_2 F_1(\tfrac{1}{2}+\tfrac{m_2-\sqrt{u}}{\hbar},\tfrac{1}{2}+\tfrac{m_2+\sqrt{u}}{\hbar},1+\tfrac{m_2-m_1}{\hbar}|-y) \label{psiy}
\end{aligned}
\end{equation}
where $d_{1,2}$ are integration constants. Let us see consider now the difference equation \eqref{difference}. For $q=0$, the difference equation reduces to
\begin{equation}
W(x) = -\frac{1}{P_0(x)}
\end{equation}
leading to
\be
\widetilde U(x+\hbar) =-{P_0(x) \over P_R(x+\ft{\hbar}{2}) } \widetilde U(x)   
\ee
The solution can be written in terms of Gamma functions as
\begin{equation}
\widetilde U (x)=
\frac{ (-1)^{\frac{x}{\hbar}}  \Gamma (\tfrac{x-\sqrt{u}}{\hbar}) \Gamma (\tfrac{x+\sqrt{u}}{\hbar})}
{\Gamma (\tfrac{1}{2}+\tfrac{x-m_1}{\hbar}) \Gamma (\frac{1}{2}+\frac{x-m_2}{\hbar})}
\end{equation}
up to an overall constant. The connection with the wave function $U(y)$ is given by the Laplace transform of $\widetilde U(x)$, defined as
\begin{equation}
U(y)=\frac{1}{2\pi i}\int_{\gamma} y^{\frac{x}{\hbar}} \,  \widetilde U(x) \, dx
\end{equation}
with $\gamma$ a contour enclosing the two series of poles: $x= \sqrt{u}-\hbar n$ and $x= -\sqrt{u}-\hbar n$. Computing the residues and summing over $n$ one finds
\begin{equation}
\frac{(-y)^{\frac{\sqrt{u}}{\hbar }}  \Gamma(\frac{2 \sqrt{u}}{\hbar})
{}_2F_1\left(\tfrac{1}{2}{+}\tfrac{m_1{-}\sqrt{u}}{\hbar},\tfrac{1}{2}{+}\tfrac{m_2{-}\sqrt{u}}{\hbar},1{-}\tfrac{2\sqrt{u}}{\hbar}|{-}y^{-1}\right)}
{\Gamma (\frac{1}{2}{+}\frac{\sqrt{u}-m_1}{\hbar}) \Gamma (\frac{1}{2}{+}\frac{\sqrt{u}-m_2}{\hbar})}
+(\sqrt{u}\to {-}\sqrt{u})
\end{equation}
that after using hypergeometric identities matches precisely (\ref{psiy}) for a constant $d_1=d_2$. 

\section{Tables}
\label{appendixTables}

In this appendix we show the QNMs of scalar perturbations of the metric with $\ell = m_\phi = 2$ and $n = 0,1$. We start with the Reissner-Nordstr{\"o}m BH, then move to the generic Kerr-Newman BH with ${\cal Q} = 0.5 \,{\cal M}$ and $a_{_\mathcal{J}}= 0.5 \,{\cal M}$. The BH mass ${\cal M}$ is always set to one.


In the tables we show the results of the computation performed using three methods: the geodesic approximation (described in section \ref{geo_approx}), then the quantization of the WKB cycle (using the SW prepotential with fourth instantons contributions, as shown in section \ref{quantum_periods_lesser_flavours}), and finally Leaver's numerical method (illustrated in section \ref{numerical_analysis}).

\subsection{Reissner-Nordstr{\"o}m BH}
$n_\text{geo} = 0$
\begin{equation}
\begin{array}{||c||c|c|c||}
\hline
 \mathcal{Q} & \omega _{\text{geo}} & \omega _{\text{SW},4} & \omega _{\text{Num}} \\
\hline
 0 & 0.471-0.0962 i & 0.476-0.0958 i & 0.484-0.0968 i \\
\hline
 0.1 & 0.472-0.0963 i & 0.477-0.0959 i & 0.484-0.0968 i \\
\hline
 0.3 & 0.479-0.0967 i & 0.484-0.0966 i & 0.491-0.0972 i \\
\hline
 0.6 & 0.504-0.0978 i & 0.511-0.0987 i & 0.517-0.0983 i \\
\hline
 0.9 & 0.567-0.0963 i & 0.581-0.106 i & 0.582-0.0966 i \\
\hline
\end{array}
\end{equation}
\\
$n_\text{geo} = 1$
\begin{equation}
\begin{array}{||c||c|c|c||}
\hline
 \mathcal{Q} & \omega _{\text{geo}} & \omega _{\text{SW},4} & \omega _{\text{Num}} \\
\hline
 0 & 0.471-0.289 i & 0.467-0.297 i & 0.464-0.296 i \\
\hline
 0.1 & 0.472-0.289 i & 0.467-0.297 i & 0.465-0.296 i \\
\hline
 0.3 & 0.479-0.29 i & 0.475-0.298 i & 0.472-0.297 i \\
\hline
 0.6 & 0.504-0.294 i & 0.502-0.299 i & 0.499-0.299 i \\
\hline
 0.9 & 0.567-0.289 i & 0.567-0.292 i & 0.567-0.293 i \\
\hline
\end{array}
\end{equation}

\subsection{Kerr-Newman BH: \texorpdfstring{${\cal Q} = 0.5 \,{\cal M}$}{Q=0.5M}}
$n_\text{geo} = 0$
\begin{equation}
\begin{gathered}
\begin{array}{||c||c|c|c||}
\hline
 a_{_\mathcal{J}} & \omega _{\text{Geo}} & \omega _{\text{SW},4} & \omega _{\text{Num}} \\
\hline
 0 & 0.493-0.0974 i & 0.497-0.105 i & 0.506-0.0979 i \\
\hline
 0.1 & 0.511-0.0972 i & 0.514-0.103 i & 0.524-0.0977 i \\
\hline
 0.2 & 0.531-0.0967 i & 0.533-0.0997 i & 0.545-0.0972 i \\
\hline
 0.3 & 0.555-0.0957 i & 0.553-0.0952 i & 0.569-0.0963 i \\
\hline
 0.4 & 0.582-0.0942 i & 0.577-0.089 i & 0.596-0.0947 i \\
\hline
\end{array}
\\
\begin{array}{||c||c|c|c||}
\hline
 a_{_\mathcal{J}} & A_{\text{geo}} & A_{\text{SW},4} & A_{\text{Num}} \\
\hline
 0 & 6. & 6. & 6. \\
\hline
 0.1 & 6.+0.000248 i & 6.+0.000142 i & 6.+0.000146 i \\
\hline
 0.2 & 6.+0.00103 i & 6.+0.000586 i & 6.+0.000605 i \\
\hline
 0.3 & 5.99+0.0024 i & 6.+0.00135 i & 6.+0.00141 i \\
\hline
 0.4 & 5.99+0.00442 i & 5.99+0.00243 i & 5.99+0.00259 i \\
\hline
\end{array}
\end{gathered}
\end{equation}
\\
$n_\text{geo} = 1$
\begin{equation}
\begin{gathered}
\begin{array}{||c||c|c|c||}
\hline
 a_{_\mathcal{J}} & \omega _{\text{Geo}} & \omega _{\text{SW},4} & \omega _{\text{Num}} \\
\hline
 0 & 0.493-0.292 i & 0.49-0.299 i & 0.487-0.299 i \\
\hline
 0.1 & 0.511-0.292 i & 0.51-0.298 i & 0.507-0.298 i \\
\hline
 0.2 & 0.531-0.29 i & 0.532-0.297 i & 0.529-0.296 i \\
\hline
 0.3 & 0.555-0.287 i & 0.558-0.294 i & 0.554-0.292 i \\
\hline
 0.4 & 0.582-0.283 i & 0.588-0.289 i & 0.584-0.287 i \\
\hline
\end{array}
\\
\begin{array}{||c||c|c|c||}
\hline
 a_{_\mathcal{J}} & A_{\text{geo}} & A_{\text{SW},4} & A_{\text{Num}} \\
\hline
 0 & 6. & 6. & 6. \\
\hline
 0.1 & 6.+0.000248 i & 6.+0.000142 i & 6.+0.000146 i \\
\hline
 0.2 & 6.+0.00103 i & 6.+0.000586 i & 6.+0.000605 i \\
\hline
 0.3 & 5.99+0.0024 i & 6.+0.00135 i & 6.+0.00141 i \\
\hline
 0.4 & 5.99+0.00442 i & 5.99+0.00243 i & 5.99+0.00259 i \\
\hline
\end{array}
\end{gathered}
\end{equation}

\subsection{Kerr-Newman BH: \texorpdfstring{$a_{_{\cal J}} = 0.5 \,{\cal M}$}{a=0.5M}}
$n_\text{geo} = 0$
\begin{equation}
\begin{gathered}
\begin{array}{||c||c|c|c||}
\hline
 \mathcal{Q} & \omega _{\text{geo}} & \omega _{\text{SW},4} & \omega _{\text{Num}} \\
\hline
 0 & 0.572-0.0929 i & 0.561-0.0807 i & 0.586-0.0935 i \\
\hline
 0.1 & 0.574-0.0929 i & 0.563-0.0807 i & 0.587-0.0935 i \\
\hline
 0.2 & 0.578-0.0929 i & 0.567-0.0808 i & 0.592-0.0934 i \\
\hline
 0.3 & 0.586-0.0927 i & 0.574-0.0809 i & 0.6-0.0933 i \\
\hline
 0.4 & 0.598-0.0924 i & 0.585-0.0811 i & 0.612-0.093 i \\
\hline
\end{array}
\\
\begin{array}{||c||c|c|c||}
\hline
 \mathcal{Q} & A_{\text{geo}} & A_{\text{SW},4} & A_{\text{Num}} \\
\hline
 0 & 5.98+0.00672 i & 5.99+0.00357 i & 5.99+0.00392 i \\
\hline
 0.1 & 5.98+0.00674 i & 5.99+0.00357 i & 5.99+0.00393 i \\
\hline
 0.2 & 5.98+0.00679 i & 5.99+0.00359 i & 5.99+0.00396 i \\
\hline
 0.3 & 5.98+0.00687 i & 5.99+0.00363 i & 5.99+0.00401 i \\
\hline
 0.4 & 5.98+0.00699 i & 5.99+0.00367 i & 5.99+0.00408 i \\
\hline
\end{array}
\end{gathered}
\end{equation}
\\
$n_\text{geo} = 1$
\begin{equation}
\begin{gathered}
\begin{array}{||c||c|c|c||}
\hline
 \mathcal{Q} & \omega _{\text{geo}} & \omega _{\text{SW},4} & \omega _{\text{Num}} \\
\hline
  0 & 0.572-0.279 i & 0.578-0.287 i & 0.573-0.283 i \\
\hline
 0.1 & 0.574-0.279 i & 0.579-0.287 i & 0.575-0.283 i \\
\hline
 0.2 & 0.578-0.279 i & 0.584-0.287 i & 0.58-0.283 i \\
\hline
 0.3 & 0.586-0.278 i & 0.593-0.286 i & 0.588-0.283 i \\
\hline
 0.4 & 0.598-0.277 i & 0.606-0.285 i & 0.601-0.281 i \\
\hline
\end{array}
\\
\begin{array}{||c||c|c|c||}
\hline
 \mathcal{Q} & A_{\text{geo}} & A_{\text{SW},4} & A_{\text{Num}} \\
\hline
 0 & 5.98+0.0202 i & 5.99+0.011 i & 5.99+0.0116 i \\
\hline
 0.1 & 5.98+0.0202 i & 5.99+0.0111 i & 5.99+0.0117 i \\
\hline
 0.2 & 5.98+0.0204 i & 5.99+0.0111 i & 5.99+0.0117 i \\
\hline
 0.3 & 5.98+0.0206 i & 5.99+0.0112 i & 5.99+0.0119 i \\
\hline
 0.4 & 5.98+0.021 i & 5.99+0.0114 i & 5.99+0.0121 i \\
\hline
\end{array}
\end{gathered}
\end{equation}

\end{appendix}
\newpage

\bibliographystyle{JHEP}
\bibliography{Ref}
  
\end{document}